\documentclass[12pt,preprint]{aastex}
\usepackage{amsmath,euscript,amsfonts}

\def\simlt{\mathrel{\hbox{\rlap{\hbox{\lower4pt\hbox{$\sim$}}}\hbox{$<$}}}}
\def\simgt{\mathrel{\hbox{\rlap{\hbox{\lower4pt\hbox{$\sim$}}}\hbox{$>$}}}}

\def\ale{\mathrel{\hbox{\rlap{\hbox{\lower4pt\hbox{$\sim$}}}\hbox{$<$}}}}
\def\age{\mathrel{\hbox{\rlap{\hbox{\lower4pt\hbox{$\sim$}}}\hbox{$>$}}}}

\def\nodata{---}

\def\cit{1}
\def\ociw{2}
\def\pu{3}
\def\hf{4}
\def\uva{5}
\def\vla{6}

\shorttitle{The Radio and X-ray Luminous Type Ibc Supernova 2003L}
\shortauthors{Soderberg et al.}

\begin{document}

\title{The Radio and X-ray Luminous Type Ibc Supernova 2003L}

\author{
A.~M. Soderberg\altaffilmark{\cit}, 
S.~R. Kulkarni\altaffilmark{\cit}, 
E. Berger\altaffilmark{\ociw,\pu,\hf},
R.~A. Chevalier\altaffilmark{\uva},
D.~A. Frail\altaffilmark{\vla},
D.~B. Fox\altaffilmark{\cit},
R.~C. Walker\altaffilmark{\vla}
}

\altaffiltext{\cit}{Division of Physics, Mathematics and Astronomy,
        105-24, California Institute of Technology, Pasadena, CA
        91125}
\altaffiltext{\ociw}{Observatories of the Carnegie Institution of Washington,
        813 Santa Barbara St., Pasadena, CA 91101}
\altaffiltext{\pu}{Department of Astrophysical Sciences, Princeton University, 
        Princeton, NJ 08544}
\altaffiltext{\hf}{Hubble Fellow}
\altaffiltext{\uva}{Department of Astronomy, University of Virginia, P.O. Box 3818, Charlottesville, VA 22903-0818}
\altaffiltext{\vla}{National Radio Astronomy Observatory, Socorro,
        NM 87801}

\begin{abstract}
We present extensive radio observations of SN\,2003L, the most
luminous and energetic Type Ibc radio supernova with the exception of
SN\,1998bw (associated with GRB\,980425).  Observations from the Very
Large Array are well described by a fitting a synchrotron
self-absorption model to the emission spectrum.  This model implies a
sub-relativistic ejecta velocity, $\overline{v}\approx 0.2 c$, and a
size of $r\approx 4.3\times 10^{15}$ cm at $t\approx 10$ days.  The
circumstellar density is suitably fit with a stellar wind profile,
$n_e \propto r^{-2}$ and a constant mass loss rate of
$\dot{M}\approx 7.5 \times 10^{-6}~\rm M_{\odot}~yr^{-1}$.  Moreover,
the magnetic field follows $B\propto r^{-1}$ and the
kinetic energy of the radio bright ejecta is roughly $E\approx
10^{48}$ erg assuming equipartition of energy between relativistic
electrons and magnetic fields. Furthermore, we show that free-free
absorption does not contribute significantly to the radio spectrum,
since it implies ejecta velocities which are inconsistent with size
constraints derived from Very Long Baseline Array observations.  In
conclusion, we find that although SN\,2003L has a radio luminosity
comparable to that seen in SN\,1998bw, it shows no evidence for a
significant amount of energy coupled to relativistic ejecta.  Using
SN\,2003L as an example, we comment briefly on the coupling of ejecta
velocity and energy in Type Ibc supernovae.
\end{abstract}


\section{Introduction}
Radio emission from core-collapse supernovae (SNe) have been sought
primarily to study the circumstellar medium
\citep{wsp+86,c98}. However, starting with Type Ibc SN\,1998bw,
associated with the under-luminous gamma-ray burst GRB\,980425
\citep{paa+00,gvv+98,kfw+98}, radio observations have also been
used a tracer of relativistic ejecta \citep{kfw+98,lc99}.  This event
prompted a storm of publications: SN\,1998bw was the most luminous
radio supernova (at early times), showed strong variability and
produced copious gamma-ray emission relative to ordinary SNe Ibc.
These data provided the first observational evidence that at least
some SNe Ibc are associated with GRBs
(e.g.~\citealt{bkd+99,bkp+02,smg+03}, c.f.~\citealt{pks+03}).  

To further explore the connection between GRBs and SNe Ibc, we 
began a radio survey of local Type Ibc supernovae. Our survey was motivated
by two simple phenomenological questions: (1) What is the prevalence of
SN\,1998bw-like supernovae?, and (2) Is there a continuum between ordinary 
SNe and GRBs as traced by their relativistic ejecta?

In \citet{bkf+03} we reported the first three years of the survey.  We
found that most SNe Ibc are not detectable with current radio
sensitivity, with limits reaching $10^{-3}$ that of the peak
brightness of SN\,1998bw.  Clearly, luminous SN\,1998bw-like supernovae
are rare (less than 3\%) and those within an order of magnitude
of SN\,1998bw are uncommon (less than 10\%).

Here, we present radio observations of SN\,2003L, the first SN Ibc
with a radio luminosity comparable to SN\,1998bw.  While there is no
evidence for mildly-relativistic ejecta, the radio emitting ejecta of
SN\,2003L are unusually energetic.  This discovery suggests the
existence of a sub-class of SNe Ibc with similar energetics to
SN\,1998bw and even weaker central engines, if any.

The organization of this paper is as follows: observations from the
Very Large Array (VLA), the Very Long Baseline Array (VLBA) and the
{\it Chandra} X-ray Observatory (CXO) are described in \S\ref{sec:obs}
and \S\ref{sec:xray}.  Preliminary estimates of the energy and
velocity are presented in \S\ref{sec:ep}.  Modeling of the radio
light-curves is discussed in Sections~\ref{sec:ssa} and \ref{sec:ffa}
assuming the dominant absorption is from internal
synchrotron self-absorption and external free-free absorption, respectively,
Modeling of the X-ray emission follows as \S\ref{sec:xray_model}.
In the Discussion (\S\ref{sec:disc}) we conclude by addressing the
nature of SN\,2003L.

\section{Radio Observations of SN\,2003L}
\label{sec:obs}
\subsection{Very Large Array Data}

SN\,2003L was optically discovered on 2003 January 12.15 UT, offset 9''.0 W
and 1".5 N from the center of NGC 3506 \citep{b03} and at a distance of
$d\approx 92$ Mpc \citep{vcd+03}.  In our first observation with the
Very Large Array\footnote{The Very Large Array and Very Long Baseline
Array are operated by the National Radio Astronomy Observatory, a
facility of the National Science Foundation operated under cooperative
agreement by Associated Universities, Inc.} (VLA), on 2003 January
26.2 UT, we detected a radio source coincident with the optical
position at $\alpha\rm (J2000)=11^{\rm h}03^{\rm m}12.3^{\rm s},
\delta\rm (J2000)=+11^{\rm o}04'38.1''$ ($\pm$0.1 arcsec in each
coordinate) with flux density of $f_{\nu}=743\pm 39~\mu$Jy at 8.5 GHz.
This, and subsequent observations at 4.9, 8.5, 15.0 and 22.5 GHz, are
summarized in Table~1.

All observations were taken in standard continuum observing mode with
a bandwidth of $2\times 50$ MHz.  We used 3C286 for flux calibration,
and for phase referencing we used calibrators J1118+125, J1120+134 and
J1103+119.  Data were reduced using standard packages within the
Astronomical Image Processing System (AIPS).

\begin{figure}
\plotone{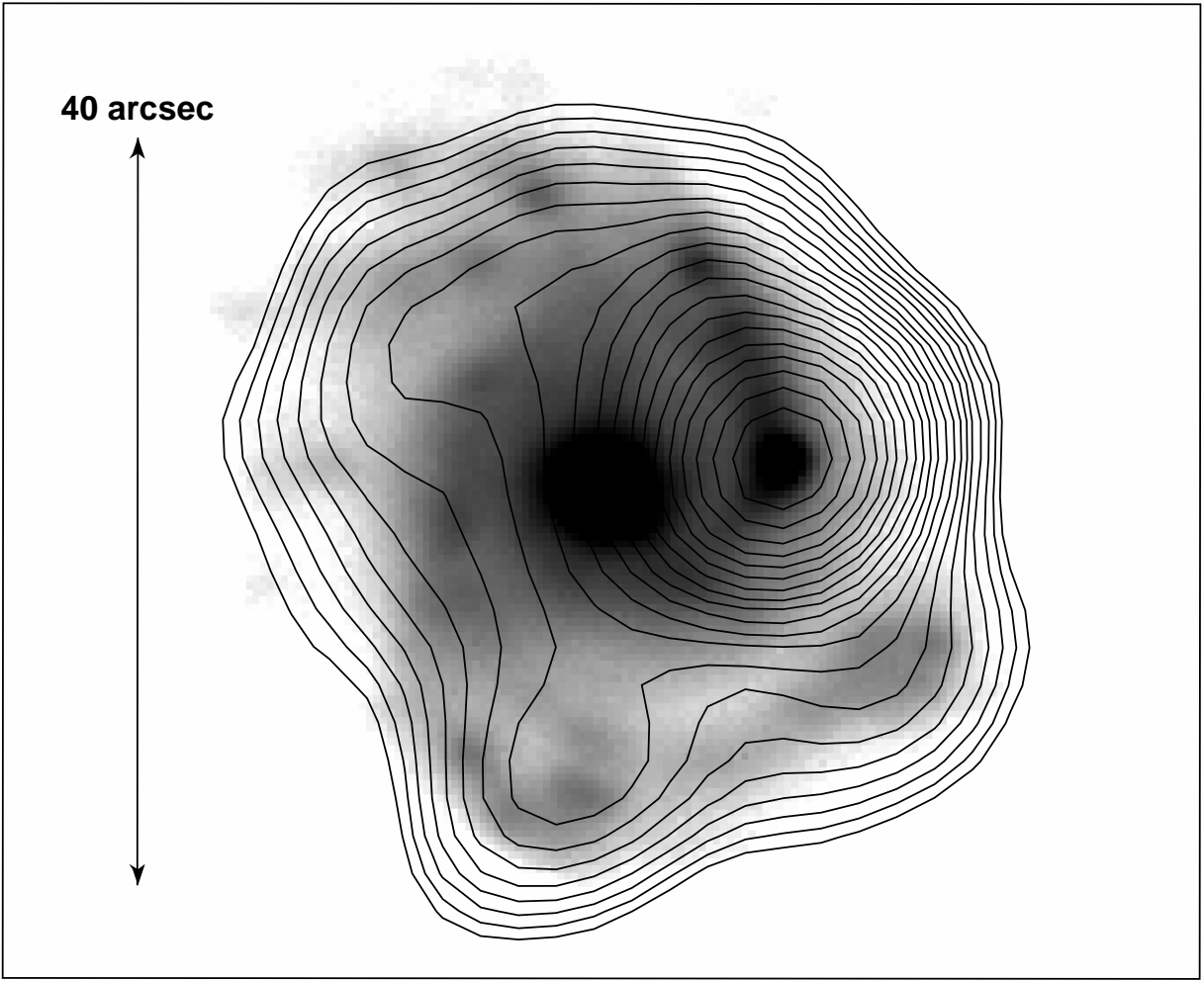}
\caption{Optical and radio emission from SN\,2003L at $t\sim 35$ days.
A Palomar 60-inch $R$-band image (greyscale) shows the SN located 
within a spiral arm of host galaxy NGC 3506.  Twenty-five radio (8.5
GHz) map contours are over-plotted from 0.15 mJy (3-sigma) to 4.0 mJy
in equally spaced logarithmic intervals.  The radio emission peaks at the
optical position of SN\,2003L and diffuse emission from
the host galaxy extends a size of approximately 45 by 45 arcsec.
\label{fig:SN_contours}}
\end{figure}

As seen in Table~1, observations were taken in each of the
four VLA configurations.  The decreased spatial resolution of the C
and D configurations resulted in detection of diffuse host galaxy
emission surrounding the supernova, primarily affecting the low ($\le
8.5$ GHz) frequency observations (Figure~\ref{fig:SN_contours}).  At
8.5 GHz the host galaxy extends $45\times 45$ arcsec, but is resolved
away on baselines $\geq 10~\rm k\lambda$.  By restricting the UV range
to exclude shorter baselines, we measured the flux density of the
supernova with minimal contribution from the host galaxy.
Measurements at higher frequencies and in other VLA configurations were
made by fitting a Gaussian to the SN emission and solving for
the integrated intensity.

\begin{figure}
\plotone{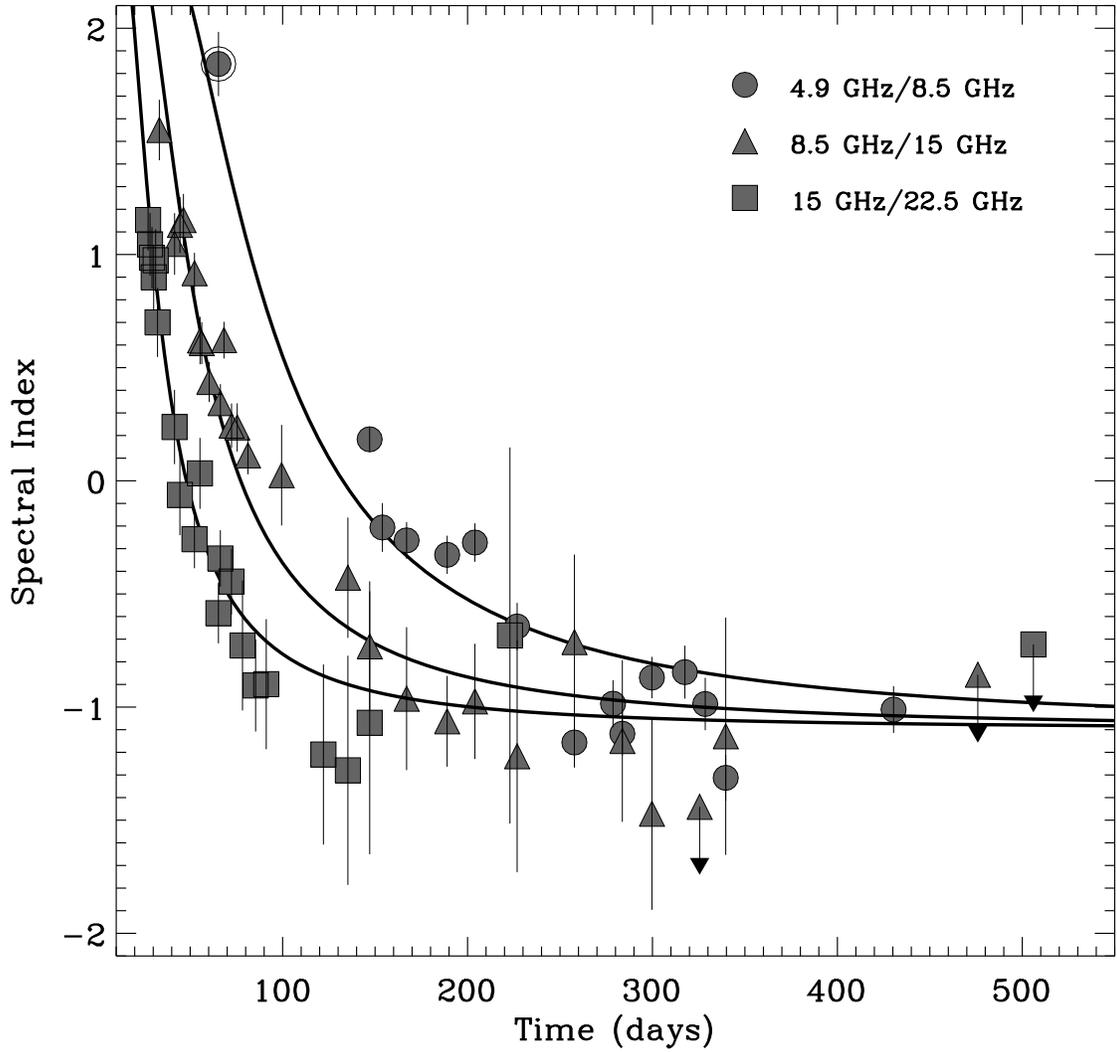}
\caption{Spectral indices for the radio emission of SN 2003L.  The
temporal evolution of 4.9/8.5 GHz, 8.5/15.0 GHz, and 15.0/22.5 GHz
spectral indices are shown and SSA Model 1 (\S\ref{sec:ssa}) is
over-plotted.  We highlight the VLBA observation with an encircled
dot.
\label{fig:indices}}
\end{figure}

Figure~\ref{fig:lt_curves} shows the SN\,2003L radio light-curves
assuming an approximate explosion date of 2003 January 1.0 UT as
derived from optical data \citep{skg+04}.  The radio light-curves evolve
as $f_{\nu}\propto t^{1.2}$ before the peak, exhibit a broad maximum
and fade as $f_{\nu}\propto t^{-1.2}$.  The spectral index ($\beta$
with $f_{\nu}\propto \nu^{\beta}$) is observed as $\beta \approx -1.1$
in the optically thin regime (Figure~\ref{fig:indices}).

\subsection{Very Long Baseline Array Data}
\label{sec:vlba}
Complimentary to our VLA campaign, we obtained a single observation of
SN\,2003L using the Very Large Baseline Array (VLBA).  The six hour
observation began on 2003 March 7.30 UT ($t\approx 65$ days), and was
taken in standard continuum mode with a bandwidth of $4\times 8$ MHz
centered on the observing frequency of 4.9 GHz.  Fringe calibrations
were applied using 3C273 and phase referencing was conducted
using J1103+1158 at an angular distance of 0.9 degrees from the SN.

\begin{figure}
\plotone{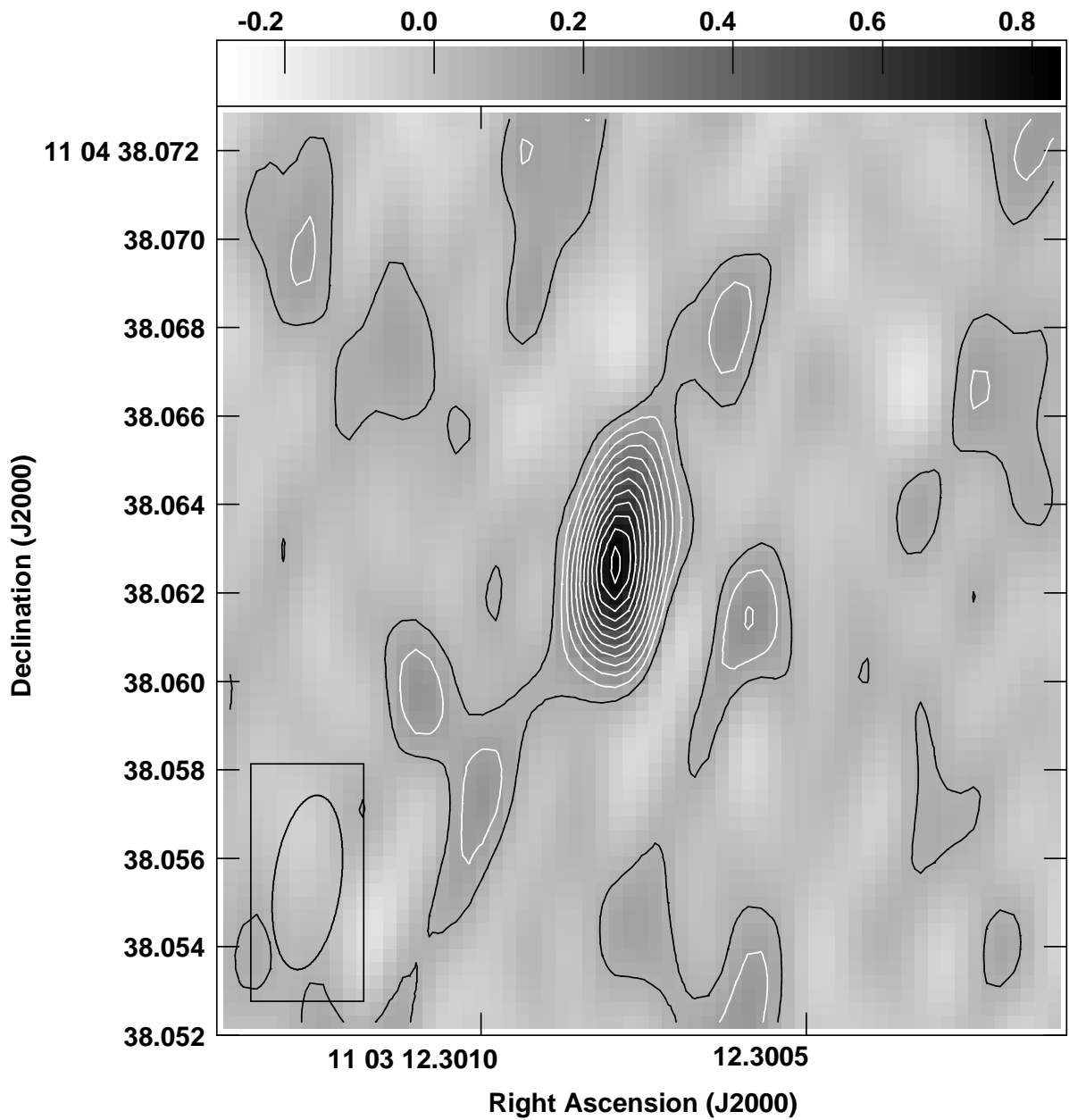}
\caption{VLBA image of SN2003L at 4.9 GHz at $t\approx 65$ days after
the explosion.  We find a flux density for the supernova of
$F_{4.9\rm~GHz}=848 \pm 64~\mu\rm Jy$ within a 3.158 x 1.285 mas
beam. Contours define 1 $\sigma$ increments of $67~\mu\rm~Jy$. 
At a distance of 92 Mpc, this unresolved VLBA
detection places a direct constraint on the size of the expanding
ejecta of $r\lesssim 9.1\times 10^{17}$ cm at $t\approx 65$ days.
\label{fig:vlba}}
\end{figure}

We detect SN\,2003L at a position of $\alpha=11^{\rm h} 03^{\rm m}
12^{\rm s}.3008$, $\delta=+11^{\circ} 04' 38''.08$ (J2000.0) with a
positional uncertainty of 10 mas in each coordinate
(Figure~\ref{fig:vlba}).  We note that these errors are dominated by
the positional uncertainty of J1103+1158, assumed to be at
$\alpha=11^{\rm h} 03^{\rm m} 03^{\rm s}.5299$, $\delta=+11^{\circ}
58' 16''.61$ (J2000.0).  Using the VLBA utilities within AIPS, we find
a flux density for the supernova of $F_{4.9\rm~GHz}=848 \pm 64~\mu$Jy
within the $3.158\times 1.285$ mas beam. We note that there is no
emission from the host galaxy at this resolution.  At a distance of 92
Mpc, this unresolved VLBA detection places a direct constraint on the
size of the expanding ejecta of $r\lesssim 9.1\times 10^{17}$ cm and
an average expansion speed of $\overline{v} \le 5.4c$ during the first
65 days after the explosion.

\section{X-ray Observations with \it Chandra\rm}
\label{sec:xray}

\begin{figure}
\plotone{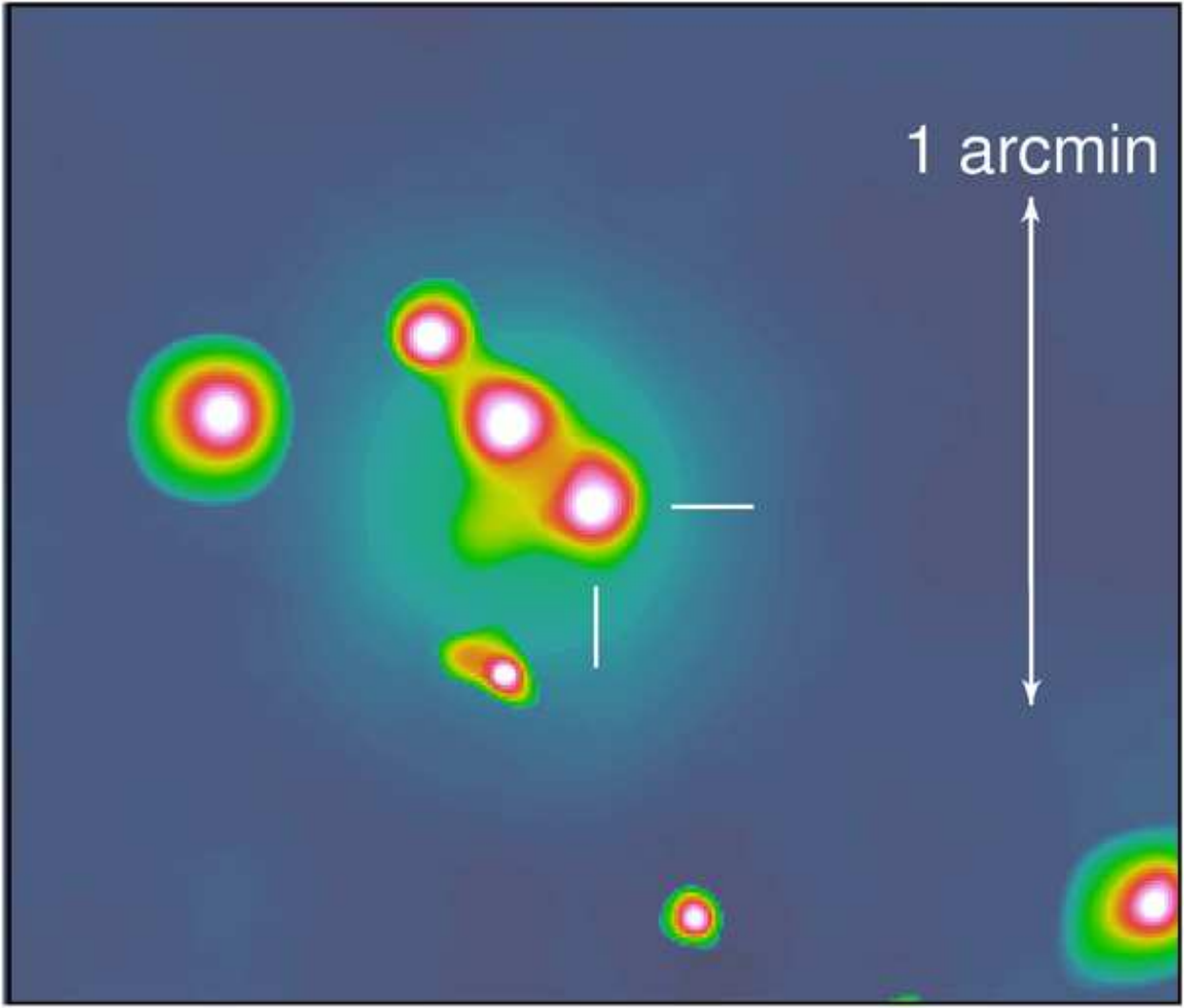}
\bigskip
\bigskip
\caption{{\it Chandra} X-ray image of SN\,2003L at $t\approx 40$ days.
The 2--10 keV luminosity of SN\,2003L is $L_X\approx 9.2\times
10^{39}$ erg/s/Hz for a power-law spectral model.  SN\,2003L is
therefore the most luminous X-ray SN Ibc (on this timescale) with the
exception of SN\,1998bw.
\label{fig:cxo}}
\end{figure}

We observed SN\,2003L with the {\it Chandra} ACIS-S detector beginning
at 2003 February 10.15 UT ($t\approx 40$ days).  During the 30 ksec
observation, we detected $30\pm 6$ counts from the supernova
\citep{kf03}.  For spectral extraction we adopt a source aperture of
2.46~arcsec, which encloses approximately 98\% of the flux of an
on-axis point-source at 1~keV (the peak of our raw photon counts
spectrum).  Our aperture correction is therefore minimal by comparison
to our $\sim$20\% flux uncertainties (see below).  Background
counts are selected from an annular region extending from 2.46~arcsec
to 7.38~arcsec from the source; in general larger background regions
are preferred for {\it Chandra} data analysis but the extended
emission from the host galaxy makes a more local selection preferable
in this case (see Figure~\ref{fig:cxo}).

\citet{sfd+98} dust maps give $E(B-V)=0.021$\,mag for the location
of SN\,2003L.  \citet{dl90} suggest an average $N_{\rm H}$ of
$2.25\times 10^{20}\rm cm^{-2}$, while a direct conversion from the
Schlegel et al. optical reddening suggests $1.13\times
10^{20}\rm cm^{-2}$ \citep{ps95}. We fix $N_{\rm H}$ for our fits
at $2.5\times 10^{20}\rm cm^{-2}$.  

For a power-law spectral fit we find the power-law photon index
$\Gamma=1.53\pm 0.4$ (range is 0.86 to 2.24 at 90\% confidence).  For a
thermal bremsstrahlung spectral fit we find the plasma temperature has
$kT>1.7$\,keV at 90\% confidence, with a best-fit value of 6.7\,keV
that is unconstrained from higher energies.  The relatively small number of
counts enables satisfactory fits (reduced chi-squared values of
$\chi^2_r\approx 1.3$; 5 degrees of freedom) for both models,
with the neutral hydrogen absorption applied to either a power-law or
thermal bremsstrahlung continuum.

Our source count rate is $1.0\pm 0.2$\,cts/ksec in the ACIS-S3
detector.  This corresponds to a 0.5--5.0~keV flux for our spectral
models of approximately $5.3\times 10^{-15}\,~\rm erg/cm^2/s$ in
either case.  Extrapolating to the full 2--10~keV X-ray band, the
corresponding fluxes are $9.2\times 10^{-15}\,\rm~erg/cm^2/s$ for the
power law model and $7.6\times 10^{-15}\,\rm~erg/cm^2/s$ for the
thermal bremsstrahlung model.  The associated X-ray luminosity is thus
$L_{2-10~\rm keV}\approx 9.2\times 10^{39}$ erg/s and $7.2\times
10^{39}$ erg/s for the power-law and thermal models, respectively.  In
comparison with other X-ray supernovae, SN\,2003L is among the most
luminous ever detected, only a factor of $\sim 10$ fainter than
SN\,1998bw at a comparable epoch \citep{paa+00}.

\section{Preliminary Constraints}
\label{sec:ep}

We are interested in the energy and expansion velocity of the ejecta
producing the radio emission.  As a preliminary constraint, we
estimate the brightness temperature, $T_B$, of SN 2003L and compare it
with robust constraints imposed by the Inverse Compton Catastrophe
(ICC; \citealt{kp81}).  The brightness temperature of a source with
angular radius, $\theta$ is given by
\begin{equation}
T_B = \left( \frac{c^2}{2\pi k}\right) \left( \frac {f_{\nu}}{\theta^2}\right) \nu^{-2}~~K
\end{equation}
Compact sources with $T_B \gtrsim 10^{12}\,$K cool rapidly via inverse
Compton scattering.  As an initial estimate for the physical size of
the supernova, we first assume that the optical expansion velocity of
$12000\rm ~km~s^{-1}$ \citep{mck+03} can be used as an average speed
to describe the motion of the radio bright ejecta.  Using our
approximate explosion date, we estimate the shock radius to be
$r\approx2.9\times 10^{15}$ cm at $t\approx 28$ days. For an observed
flux density of $f_{\nu}\approx 3.1$ mJy at $\nu \approx 22.5$ GHz,
the brightness temperature is $T_B\approx 6.3\times 10^{11}$ K, just
below of the inverse Compton catastrophe (ICC) limit.  This exercise
provides no evidence for relativistic ejecta, therefore suggesting
that SN\,2003L expands with a modest sub-relativistic velocity.

For radio sources dominated by synchrotron-self absorption (SSA), the
brightness temperature can be further constrained to $T_B < 4\times
10^{10}$ K. By assuming equipartition between the energy in electrons
($\epsilon_e$) and magnetic fields ($\epsilon_B$), we derive an
estimate of the velocity and a lower limit on the kinetic energy of
radio emitting material. This approach has been used extensively in
discussion of extra-galactic sources and was more recently applied to
supernovae \citep{c98,kfw+98}.  The method requires three observables
at any single epoch: the spectral peak frequency, $\nu_p$, the peak
flux density, $f_p$, and the spectral index of the optically thin
emission, $f_{\nu}\propto \nu^{\beta}$.  Here, $\beta\equiv -(p-1)/2$
for electrons accelerated into a power-law distribution,
$N(\gamma)\propto \gamma^{-p}$ above a minimum Lorentz factor,
$\gamma_m$.  Following \citet{kfw+98} and \citet{bkc02}, the angular
radius of the radiosphere, $\theta_{\rm ep}$, and the equipartition
energy, $E_{\rm ep}$, are given by

\begin{equation}
\theta_{\rm ep}\approx 120 \left(\frac{d}{\rm Mpc}\right)^{-1/17} \left(\frac{f_p}{\rm mJy}\right)^{8/17} \left(\frac{\nu_p}{\rm 1~GHz}\right)^{(-2\beta-35)/34}~~\rm \mu as 
\label{eqn:theta_ep}
\end{equation}

\begin{equation}
E_{\rm ep}\approx 1.1\times 10^{56} \left(\frac{d}{\rm Mpc}\right)^{2} \left(\frac{f_p}{\rm mJy}\right)^{4} \left(\frac{\nu_p}{\rm 1~GHz}\right)^{-7} \left(\frac{\theta_{\rm ep}}{\rm \mu as}\right)^{-6}~~\rm erg. 
\label{eqn:energy_ep}
\end{equation} 

At $t\approx 30$ days (our first epoch) the peak flux density is
$f_p\approx 3.2$ mJy at $\nu_p\approx 22.5$ GHz.  Using the observed
optically thin spectral index of $\beta\approx -1.1$,
Equations~\ref{eqn:theta_ep} and \ref{eqn:energy_ep} give $\theta_{\rm
ep}\approx 7.7~\mu$as implying an equipartition radius of $r_{\rm
ep}\approx 1.1\times 10^{16}$ cm and an average expansion speed of
$\overline{v}_{\rm ep}\approx 0.15c$.  The minimum energy of the
ejecta is thus $E_{\rm ep}\approx 1.5\times 10^{47}$ erg and the
magnetic field is $B_{\rm ep}\approx 1.1$ G. These constraints alone
demand that SN\,2003L is among the most energetic Type Ibc supernovae
in terms of radio-emitting ejecta, second only to SN\,1998bw.  We note
that additional absorption terms (e.g. free-free absorption) and the
inclusion of shocked protons serve to increase the equipartition
energy estimate.

Equipartition analyses may also be used to roughly constrain the
minimum electron Lorentz factor and the characteristic synchrotron
frequency, $\nu_m$.  Equating particle kinetic energy across the shock
discontinuity, we find

\begin{equation}
\epsilon_e \left( \frac{v}{c} \right)^2 m_p c^2 \approx \overline{\gamma} m_e c^2
\end{equation}

\noindent
where $\overline{\gamma}$ is the average Lorentz factor of the
electrons and $v$ is the velocity of the ejecta.  Assuming roughly
$\overline{\gamma}\approx \gamma_m$ and substituting Equation~\ref{eqn:nu_crit}
for $\gamma_m$, we derive the following estimate for the characteristic
synchrotron frequency, $\nu_m$,

\begin{equation}
\nu_m\approx \epsilon_e^2 \left(\frac{v}{c}\right)^4 \left(\frac{m_p}{m_e}\right)^2 \left(\frac{e B}{2\pi m_e c}\right)~\rm Hz.
\end{equation}

\noindent
Adopting the values derived from equipartition analyses ($v\approx
0.15c$, $B=1.1$ G and $\epsilon_e=0.5$), we find roughly $\nu_m\approx
1$ GHz at $t\approx 30$ days which is below our observing
band.  

\section{Internal Synchrotron Self-Absorption}
\label{sec:ssa}

In the Appendix we present a rigorous formulation of the temporal and
spectral evolution of synchrotron emission arising from
sub-relativistic supernova ejecta. This prescription, based on the
formalism of \citet{fwk00}, is generalized to include the cases where
$\nu_m$, is greater than the self-absorption frequency, $\nu_a$, and
where $\epsilon_e \neq \epsilon_B$.

The observed flux density and radio spectrum at any single epoch are
determined by three parameters: $C_f$, $C_{\tau}$ and $\nu_{m,0}$.
Here, $C_f$ and $C_{\tau}$ are normalization constants of the flux
density and optical depth, respectively, while $\nu_{m,0}$ is the
value of $\nu_m$ at epoch $t_0$.  The parameters $C_f$, $C_{\tau}$ and
$\nu_{m,0}$ are in turn determined by the values of four physical
parameters at $t=t_0$: the magnetic field, $B_0$, the shock radius,
$r_0$, the minimum electron Lorentz factor, $\gamma_{m,0}$, and the
ratio of energy densities, $\frak{F}_0\equiv\epsilon_e/\epsilon_B$.
With four physical parameters ($B_0$, $r_0$, $\gamma_{m,0}$,
$\frak{F}_0$) and only three constraints ($C_f$, $C_{\tau}$,
$\nu_{m,0}$), we must assume an additional constraint in order to find
a unique solution.  An additional constraint is obtained by
adopting a value for $\frak{F}_0$. Inverting the equations for $C_f$,
$C_{\tau}$ and $\nu_{m,0}$ (Equations~\ref{eqn:C_f},~\ref{eqn:C_tau}
and \ref{eqn:nu_m_0}, respectively) we derive the following
expressions for $B_0$, $r_0$ and $\gamma_{m,0}$.

\begin{gather}
B_0=\left[1.3\times 10^8 \pi^{15} ~(2+p)^{-6}(p-2)^{-4} \frac{m_e^{19} c^{21}}{e^{17}} \frac{\eta^4}{\frak{F}_0 d^4}\right]^{1/17} \times ~C_f^{-2/17} C_{\tau}^{4/17} \nu_{m,0}^{-2(p-2)/17}~~\rm G
\label{eqn:B_0} \\
r_0=\left[1.5\times 10^{-2} \pi^{-9} ~(2+p)^7 (p-2)^{-1}  \frac{c}{m_e^8} \frac{\eta d^{16}}{\frak{F}_0}\right]^{1/17} \times ~C_f^{8/17} C_{\tau}^{1/17} \nu_{m,0}^{-(p-2)/34}~~\rm cm
\label{eqn:r_0}\\
\gamma_{m,0}=\left[3.1\times 10^{-2} \pi~\frac{(2+p)^3 (p-2)^2}{m_e c^2} \frac{\frak{F}_0^2 d^2}{\eta^2}\right]^{1/17} \times ~C_f^{1/17} C_{\tau}^{-2/17} \nu_{m,0}^{(13+2p)/34}
\label{eqn:gamma_m_0}
\end{gather}

\noindent
Here, we define $d$ as the distance to the supernova and $\eta$
characterizes the the thickness of the radiating electron shell as
$r/\eta$.  We make the standard assumption of a thin shell with $\eta
=10$ \citep{lc99,fwk00}.

\subsection{Hydrodynamical Evolution of the Ejecta}
\label{sec:hydro}
Evolutionary models governing the hydrodynamics of the SN ejecta
provide additional constraints.  As discussed by \citet{c96}, there
are several different models that can be used to describe the
hydrodynamic evolution of the ejecta.  These models allow the temporal
indices, $\alpha_r$, $\alpha_B$, $\alpha_{\gamma}$ and
$\alpha_{\frak{F}}$ to be constrained.  Here, we adopt the standard
model of \citealt{c96} for our basic SSA fit.

The first assumption of the standard model is that the hydrodynamic
evolution of the ejecta is self-similar across the shock
discontinuity.  This implies $r\propto t^{\alpha_r}$ with
$\alpha_r=(n-3)/(n-s)$ where $n$ is the density profile of the outer
SN ejecta ($\rho \propto r^{-n}$) and $s$ is the density profile of
the radiating electrons within the shocked circumstellar medium ($n_e
\propto r^{-s}$).  In addition, the standard model assumes that the
magnetic energy density ($U_B \propto B^2$) and the relativistic
electron energy density ($U_e \propto n_e \gamma_m$) scale as the
total post-shock energy density ($U \propto n_e v^2$).  In this
scenario, the magnetic field is amplified by turbulence near the shock
discontinuity, implying fixed energy fractions, $\epsilon_e$ and
$\epsilon_B$, and thus a constant value of $\frak{F}$ throughout the
evolution of the ejecta. 

Adopting this evolutionary model, the temporal indices are constrained
as follows.  The self-similar solution requires that the total energy
density scales as $U \propto t^{\alpha_{n_e} + 2\alpha_v}$ where
$\alpha_{n_e}$ defines the temporal evolution of the electron density.
For a radial dependence of $s$, we have
$\alpha_{n_e}=-s\alpha_r$. Using the scaling for the shock velocity
(Equation~\ref{eqn:v_scaling}), we then have $U \propto t^{-s
\alpha_r+2(\alpha_r-1)}$.  Similarly, $U_e \propto t^{-s
\alpha_r+\alpha_{\gamma}}$.  In a model where $\epsilon_e$ and
$\epsilon_B$ are constant, it follows that $\alpha_{\frak{F}}=0$ and
$U_e \propto U_B \propto U$.  This results in the following constraints,

\begin{gather}
\alpha_{\gamma}=2(\alpha_r-1)
\label{eqn:alpha_gamma} \\
\alpha_B=\frac{(2-s)}{2}\alpha_r - 1.
\label{eqn:alpha_B}
\end{gather}

\noindent
With the standard assumption that the circumstellar medium is
characterized by a wind density profile, $s=2$,
Equation~\ref{eqn:alpha_B} simplifies to $\alpha_B=-1$.

\subsection{SSA Model Fit for SN\,2003L}
\label{sec:ssa_model_fit}
Using the SN\,2003L multi-frequency radio light-curves and a chosen
value of $\frak{F}_0$, we fit for the constants $C_f$ and $C_{\tau}$
as well as the temporal index $\alpha_r$.  As discussed in
Section~\ref{sec:ep}, $\nu_m$ is estimated to be below our radio
observing band and we therefore adopt $\nu_{m,0}\approx 1$ GHz.  We
note that the uncertainty in $\nu_{m,0}$ corresponds to minimal
uncertainty in the derived physical parameters as shown by
Equations~\ref{eqn:B_0} -- \ref{eqn:gamma_m_0}.

\begin{figure}
\plotone{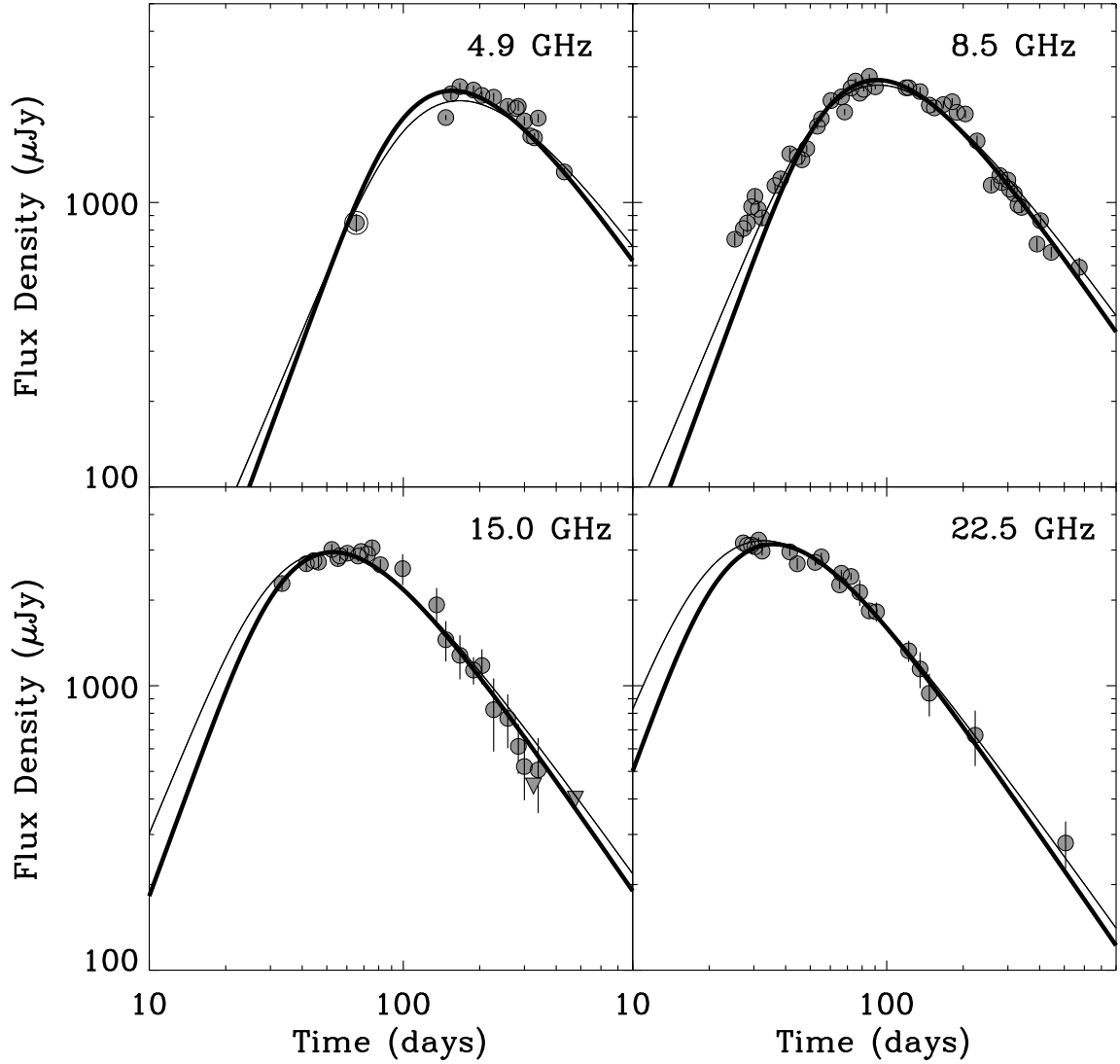}
\caption{Radio light-curves of for the Type Ic SN\,2003L were taken
  with the VLA at frequencies 4.9, 8.5, 15.0, and 22.5 GHz between
  January 2003 and July 2004.  The single VLBA observation (5 GHz) is
  shown as an encircled dot.  SSA Model 1 (thick solid line), and
  Models 2 and 3 (thin solid line) as described in \S\ref{sec:ssa}
  are over-plotted.  
\label{fig:lt_curves}}
\end{figure}

Adopting a reference time of $t_0=10$ days, we find a best-fit
solution ($\chi^2_r=7.5$; 105 degrees of freedom) for
parameters: $C_f=7.2\times 10^{-53}$, $C_{\tau}=4.5\times 10^{38}$,
and $\alpha_r=0.96$.  According to Equations~\ref{eqn:alpha_gamma} --
\ref{eqn:alpha_B}, these values imply $\alpha_{\gamma}=0.075$
and $\alpha_B=-1.0$.  Here, we have used $\frak{F}_0=1$ with
$\alpha_{\frak{F}}=0$ and an electron energy index,
$p=(-2\beta+1)=3.2$, based on the observed optically thin spectral
index, $\beta\approx -1.1$ (see Section~\ref{sec:obs} and
Figure~\ref{fig:indices}).  As discussed in the Appendix, we
parameterize the sharpness of the $\nu_a$ spectral break with
$\zeta=[0,1]$ and find $\zeta\approx 0.5$ for our best-fit solution.
The model provides a reasonable fit to the light-curves as seen in
Figure~\ref{fig:lt_curves}.  We note that the large $\chi^2_r$ is
dominated by scintillation, primarily affecting the lower frequency
observations.

Better fits can be obtained by relaxing the assumptions of the
standard model (hereafter Model 1) as described in
Section~\ref{sec:hydro}.  In Model 2, we remove the assumption of a
wind density profile and effectively fit for $s$.  For this model we
find a best-fit solution ($\chi^2_r=5.7$; 104 degrees of freedom) for
$C_f=1.2\times 10^{-52}$, $C_{\tau}=2.5\times 10^{38}$,
$\alpha_r=0.85$ and $\alpha_B=-0.84$.  By Equation~\ref{eqn:alpha_B},
this solution implies a shallow CSM density profile of $s=1.6$.  As in
the case of Model 1, we have adopted $\frak{F}_0=1$,
$\alpha_{\frak{F}}=0$, $p=3.2$ and $\zeta=0.5$ for this fit. 

A similar fit is obtained by fixing the density profile as $s=2$ and
removing the assumption of constant energy density fractions,
$\epsilon_e$ and $\epsilon_B$.  In this case, Model 3, the constraints
given by Equations~\ref{eqn:alpha_gamma} and \ref{eqn:alpha_B} do not
apply since $\alpha_{\frak{F}}$ is non-zero.  By definition, $\frak{F} \propto
U_e/U_B$, resulting in the constraint

\begin{equation}
\alpha_{\frak{F}}=-s\alpha_r + \alpha_{\gamma} -2\alpha_B
\label{eqn:alpha_frak} 
\end{equation}

\noindent
which gives $\alpha_{\frak{F}}=-2\alpha_r +\alpha_{\gamma}-2\alpha_B$
for $s=2$.  For Model 3, we find a best-fit solution ($\chi^2_r=5.7$;
103 degrees of freedom) for $C_f=1.2\times 10^{-52}$,
$C_{\tau}=2.4\times 10^{38}$, $\alpha_r=0.78$, $\alpha_B=-1.1$ and
$\alpha_{\gamma}=0.16$.  The implied evolution of $\frak{F}$ is given
by $\alpha_{\frak{F}}=0.88$ with $\epsilon_e \propto t^{0.61}$ and
$\epsilon_B \propto t^{-0.27}$.  Consistent with Models 1 and 2,
$\frak{F}_0=1$, $p=3.2$ and $\zeta=0.5$ were used for this fit.  

A comparison of the best-fit solutions from Models 1 - 3 are shown in
Figure~\ref{fig:lt_curves}.  It should be noted that Models 2 and 3
produce the same light-curve fit because since they both allow for one
extra degree of freedom.  While these two models provide a slightly
better representation of the SN\,2003L data, we adopt Model 1 as our
best SSA solution since these assumptions have been shown to be
consistent with the full physical model for SN\,1993J \citep{fb98},
currently the best-studied radio SN.

\subsection{Physical Parameters}
\label{sec:phys_params}
Using our solution for Model 1, we find the following values and
evolution for $B$, $r$ and $\gamma_m$.  The temporal evolution of the
magnetic field is $B\approx 4.5 (t/t_0)^{-1.0}$ G with an associated
radial dependence, $B\propto r^{-1.04}$, similar to that found for the
Type IIb SN 1993J ($B\propto r^{-1}$; \citealt{fb98}).  The shock
radius is well described by $r\approx 4.3\times 10^{15}
(t/t_0)^{0.96}$ cm implying a sub-relativistic velocity of
$\overline{v} \approx 0.20 (t/t_0)^{-0.04}~c$.  We note that this
radial evolution is faster to that of SN\,1983N ($r \propto
t^{0.86}$; \citealt{c98}) and is within the expected range, $0.67 \le
\alpha_r \le 1.0$ \citep{c96,c98}.  The minimum Lorentz factor of the
electrons follows $\gamma_m \approx 8.9 (t/t_0)^{-0.08}$ and the
implied evolution of $\nu_m$ (see Equation~\ref{eqn:nu_m}) is
$\nu_m\approx 1.0 (t/t_0)^{-1.16}$ GHz which is nearly comparable to the
evolution observed for SN\,1998bw ($\nu_m \propto t^{-1.0}$;
\citealt{lc99}) and significantly slower than the decay rate in the
adiabatic Sedov-Neumann-Taylor solution ($\nu_m \propto t^{-3}$;
\citealt{zr02,fwk00}).

Using these derived values, additional physical parameters can be
computed.  The number density of emitting electrons within the
circumstellar ejecta, $n_e$, is thus

\begin{equation}
n_e=\frac{p-2}{p-1} \frac{B_0^2}{8\pi} \frac{\frak{F}_0}{m_e c^2 \gamma_{m,0}} \left(\frac{t}{t_0}\right)^{2\alpha_B-\alpha_{\gamma}+\alpha_{\frak{F}}}~~\rm cm^{-3}.
\label{eqn:n_e}
\end{equation}

\noindent
Using our above values for $\frak{F}_0$, $B$ and $\gamma_{m,0}$, we find
$n_e\approx 6.1 \times 10^{4} (r/r_0)^{-2.0}~\rm cm^{-3}$.    For
$\alpha_r\approx 0.96$ and $s=2$, the density profile of
the ejecta scales steeply as $\rho \propto r^{-27}$.

The associated progenitor mass loss rate is 

\begin{equation}
\dot{M}=\frac{8\pi}{\eta} n_e m_p r_0^2 v_w \left(\frac{t}{t_0}\right)^{2\alpha_B-\alpha_{\gamma}+2\alpha_r+\alpha_{\frak{F}}}~~\rm M_{\odot}~yr^{-1}
\end{equation}
\label{eqn:m_dot}

\noindent
where we have assumed a nucleon-to-proton ratio of 2 and $v_w$ is the
velocity of the stellar wind.  Assuming a typical wind velocity of
$v_w=1000~\rm km~s^{-1}$, we find a progenitor mass loss rate of
$\dot{M}\approx 7.5\times 10^{-6}~M_{\odot}~\rm yr^{-1}$ for
SN\,2003L.  We note that this derived mass loss rate is a factor of
$\sim 30$ larger than that of the Type Ic SN\,1998bw and SN\,2002ap
and yet falls at the low end of values observed for Galactic
Wolf-Rayet (WR) stars \citep{cgv04}, the favored progenitor model for
Type Ibc SNe and gamma-ray bursts.

\noindent
The total ejecta energy, $E$, is then given as

\begin{equation}
E=\frac{4\pi}{\eta} r_0^3 \frac{\frak{F}_0}{\epsilon_{e,0}} \frac{B_0^2}{8\pi} \left(\frac{t}{t_0}\right)^{2\alpha_B+3\alpha_r+\alpha_{\frak{F}}-\alpha_{\epsilon_e}}~\rm erg.
\label{eqn:energy}
\end{equation}

\noindent
It should be noted that the total energy depends not only on the
ratio $\frak{F}_0$, but also on assumed value of $\epsilon_{e,0}$.
For $\epsilon_{e,0}=0.1$ (consistent with \citealt{lc99,bkc02})
we find an ejecta energy of $E_0\approx 8.2\times 10^{47}
(t/t_0)^{0.89}$ erg, just a factor of two below that of SN\,1998bw and
300 times larger than that of SN\,2002ap at similar epochs.  As
discussed by \citet{lc99}, a larger
$\frak{F}_0$ would increase the energy budget further.  

For comparison, we derive the physical parameters implied by Models 2
and 3 here.  Model 2 shows a slower temporal evolution of radius,
$r\approx 5.3\times 10^{15}~(t/t_0)^{0.85}$ cm, and thus a more rapid
decline in velocity, $\overline{v}\approx 0.20 (t/t_0)^{-0.15} c$.
The radio bright ejecta scales as $E\approx 1.0\times 10^{48}
(t/t_0)^{0.88}$ erg and the magnetic field evolves with radius as
$B\approx 3.7 (r/r_0)^{-0.99}$ G, similar to Model 1.  The main
difference of Model 2 is that the electron density shows a radial
dependence of $n_e\approx 3.7 \times 10^4 (r/r_0)^{-1.63}~\rm
cm^{-3}$, significantly shallower than that associated with a constant
circumstellar wind.  The density profile of the
ejecta is thus given by $\rho \propto r^{-10.7}$ and similar to that
found for the Type II SN\,1987A ($\rho \propto r^{-9}$;
\citealt{a88}). This implies an evolution of the mass loss rate, 
$\dot{M}\approx 6.8\times 10^{-6} (r/r_0)^{0.37}~\rm
M_{\odot}~yr^{-1}$.

\begin{figure}
\plotone{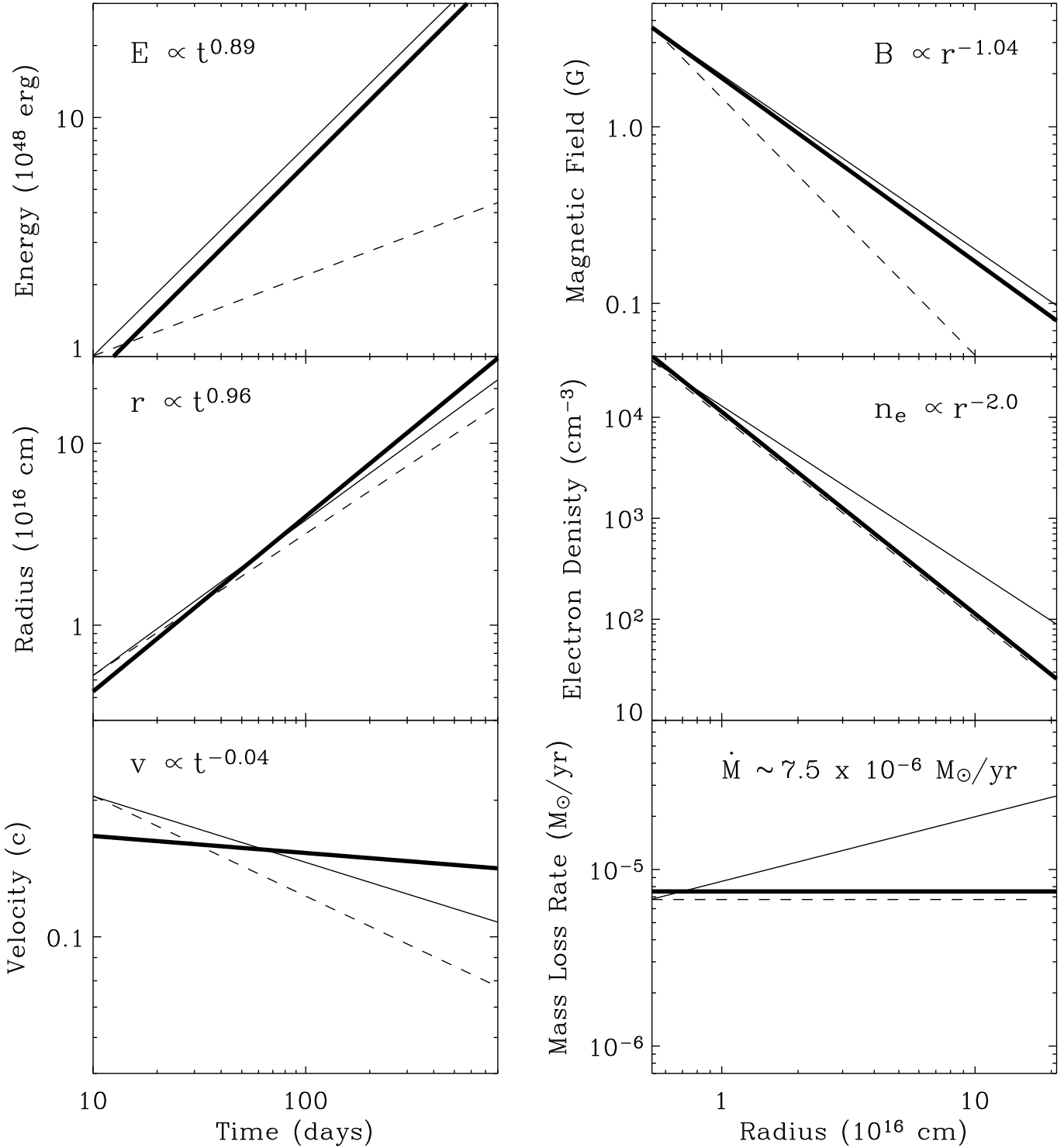}
\bigskip
\bigskip
\caption{Physical parameters for SN\,2003L based on SSA Model 1 (thick
solid line), Model 2 (thin solid line) and Model 3 (thin dashed line)
as described in Section~\ref{sec:ssa}.  Left column: the temporal
evolution from $t=10 - 800$ days is shown for the ejecta energy, shock
radius and average velocity (top to bottom).  Right column: the radial
profile of the magnetic field, electron number density and mass loss
rate are shown from $r=5\times 10^{15}$ to $2.1\times 10^{17}$ cm.
The scalings appearing at the top of each plot correspond to Model 1.
\label{fig:params}}
\end{figure}

Model 3 shows an even faster deceleration of the ejecta with
$\overline{v}\approx 0.20 (t/t_0)^{-0.22} c$ and radial evolution of
$r\approx 5.3\times 10^{15} (t/t_0)^{0.78}$ cm.  In this case, the
energy increases relatively slowly as $E\approx 1.0\times 10^{48}
(t/t_0)^{0.34}$ erg and the magnetic field falls off more steeply as
$B\approx 3.7 (r/r_0)^{-1.46}$ G.  The density profile is fixed at
$s=2$, giving an evolution of $n_e \approx 3.6 \times 10^4
(r/r_0)^{-2}~\rm cm^{-3}$ and a constant mass loss rate of
$\dot{M}\approx 6.8 \times 10^{-6}~\rm M_{\odot}~yr^{-1}$.  We note
that Model 3 predicts a relatively shallow ejecta density profile with
$\rho \propto r^{-6.25}$.  Figure~\ref{fig:params} compares the
physical parameters predicted by Models 1-3.

\section{External Absorption}
\label{sec:ffa}

In the analysis presented above we did not account for the possible effects of
external absorption on the observed radio light-curves.  For
low mass loss rates such as the value we derived through SSA modeling
of SN\,2003L (Section~\ref{sec:ssa}), it has generally been found that
free-free absorption (FFA) contributes only minimally to the observed
light-curves.  Consistent with the SSA model, FFA models assume that the
intrinsic radio emission spectrum is due to non-thermal synchrotron
processes.  The observable signature of FFA is a steepening of the
optically thick light-curves at early time (due to contributions from
both SSA and FFA processes), as was seen in the case of the Type IIb
SN\,1993J \citep{fb98}.  

\subsection{Basic FFA Model}

To begin, we fit a basic external absorption model in which the
dominant absorption process is assumed to be thermal FFA from a
uniform ionized circumstellar medium located external to the
ejecta.  Following \citep{wsp+86}, this FFA dominated model is
given by

\begin{equation}
f_{\nu} (t) = K_1 \left(\frac{\nu}{5~\rm GHz}\right)^{\beta} \left(\frac{t}{1~\rm day}\right)^{\alpha_f} e^{-\tau_{\nu}}~~\rm mJy, 
\end{equation}
\noindent
with
\begin{equation}
\tau_{\nu} (t) =K_2 \left(\frac{\nu}{5~\rm GHz}\right)^{-2.1} \left(\frac{t}{1~\rm day}\right)^{\alpha_{\tau}}.
\end{equation}

\noindent
where $\alpha_f$ and $\alpha_{\tau}$ describe the temporal evolution
of the flux density and optical depth, respectively; $K_1$ and $K_2$
are normalization constants and $\beta$ is the optically thin spectral
index.  

As shown in Figure~\ref{fig:ffa}, the basic FFA model fit (thin solid
line) provides a poor match to the data, underestimating the slow
turn-on of the radio emission with $\chi^2_r \approx 14.0$ (two times
worse than the SSA model fit, \S~\ref{sec:ssa}). The fitted parameter 
values are $K_1\approx 597$, $\beta\approx -1.0$, $\alpha_f\approx
-0.97$, $K_2\approx 1710$ and $\alpha_{\tau}\approx -1.5$.

According to this FFA dominated model, we estimate the predicted mass
loss rate of the progenitor star using Equation 17 of
\citet{wpm01}. We find a high mass loss rate of $\dot{M}\approx
1.4\times 10^{-2}~\rm M_{\odot}~yr^{-1}$ (for $v_w=1000~km~s^{-1}$).
This value is three orders of magnitude above the typical mass loss
rate for Galactic Wolf-Rayet stars, which show $\dot{M}\approx
10^{-5}~\rm M_{\odot}~yr^{-1}$ with $v_w\approx 1000$ to $2000~\rm
km~s^{-1}$ \citep{cgv04}.

\subsection{Complex FFA Model}

Better light-curve fits are obtained by applying the complex FFA model of
\citet{wpm01} which includes multiple external absorption processes,
each characterized by the distribution of absorbing material (uniform
vs clumpy) and the location of the material with respect to the
radiating electrons (within the CSM vs distant).  We note that this
FFA model also allows for both internal SSA and FFA components
to contribute to the shape of the observed radio light-curves.  

\begin{figure}
\plotone{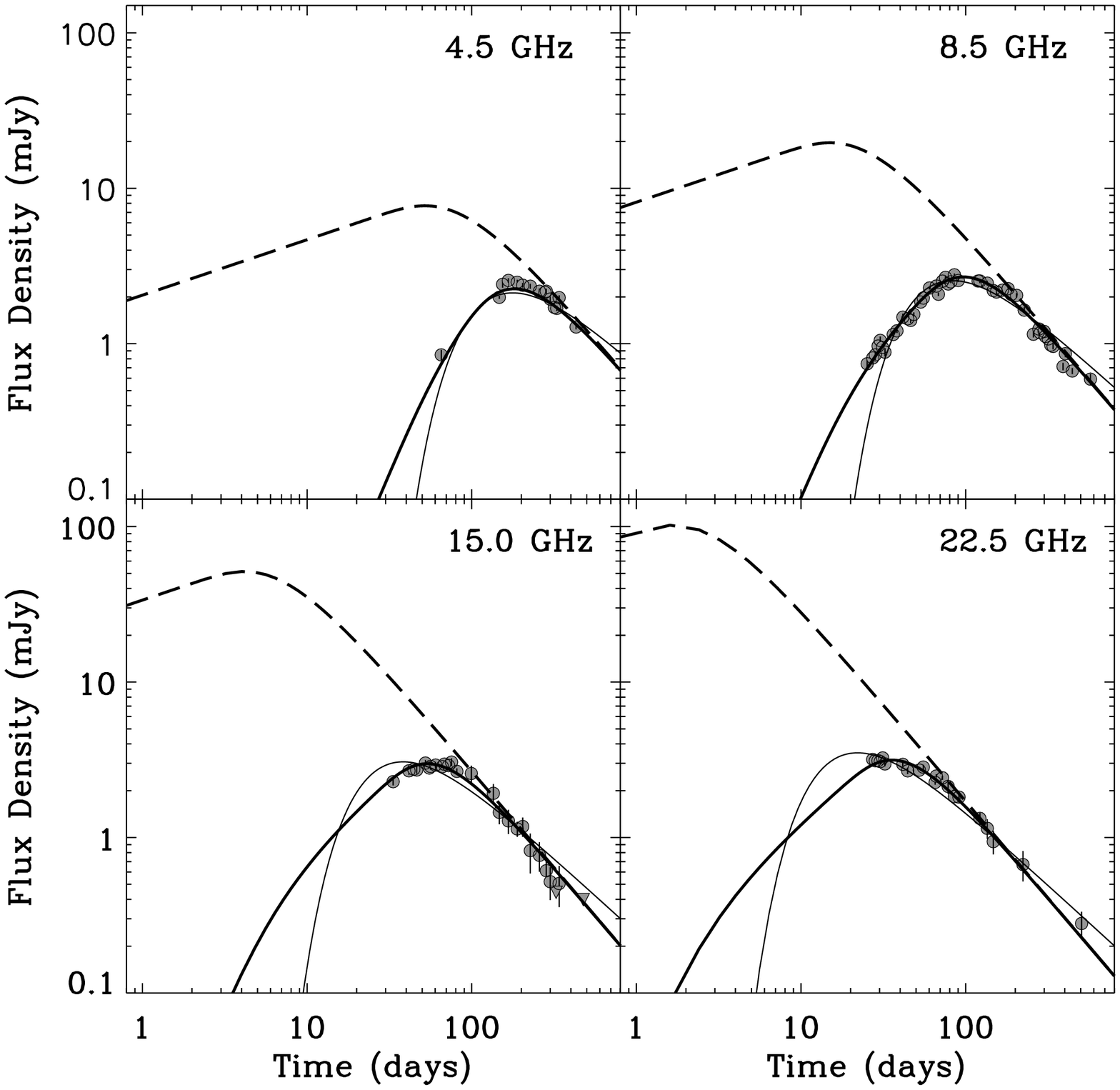}
\bigskip
\bigskip
\caption{Free-free absorption fits for SN\,2003L based on formalism
from \citet{wsp+86} (basic model; thin solid line) and \citet{wpm01}
(complex model; solid line).  The basic model provides a poor fit to
the data while the complex model provides a significantly better fit
by including additional absorption processes.  The intrinsic (de-absorbed)
synchrotron spectrum of the complex model (dashed line)
implies a relativistic velocity which violates the VLBA constraint at 
$t\approx 65$ days.
\label{fig:ffa}}
\end{figure}

Applying this model to the SN\,2003L radio light-curves, we
find that the two dominant absorption processes are internal SSA and
external FFA due to a clumpy CSM.  In the notation of \citet{wpm01}
this corresponds to $K_1\approx 2.80 \times 10^3$, $\alpha \approx
-1.11$, $\beta \approx -1.24$, $K_3 \approx 7.66\times 10^4$, $\delta
' \approx -2.15$, $K_5 \approx 1.28 \times 10^3$, $\delta '' \approx
-1.60$, with all other parameters effectively insignificant.  As shown
in Figure~\ref{fig:ffa}, the complex FFA solution (thick solid line)
provides an reasonable fit to the data with $\chi^2_r=4.5$ for 96 dof.

To estimate the physical parameters of the SN in this model we remove
the external absorption due to the clumpy CSM component leaving only
the intrinsic synchrotron self-absorbed spectrum.  As shown in
Figure~\ref{fig:ffa}, at $t\approx 2$ days the peak flux density is
$f_p\approx 102.1$ mJy at $\nu_p\approx 22.5$ GHz and $\beta \approx
-1.1$.  For these values, Equations~\ref{eqn:theta_ep} and
\ref{eqn:energy_ep} give an equipartition radius of $r_{\rm ep}\approx
5.7\times 10^{16}$ cm, an expansion velocity of $\overline{v}_{\rm
ep}\approx 14c$, and an ejecta energy of $E_{\rm ep}\approx 7.0\times
10^{48}$ erg.  With this velocity, the ejecta radius would be
$r\approx 2.4\times 10^{18}$ cm at $t\approx 65$ days, in violation of
the VLBA limit of $r\le 9.1\times 10^{17}$ cm.  To match the observed
VLBA limit, we relax the assumption of equipartition, but this results
in a steep increase in the total ejecta energy by a factor of $\sim
300$, giving $E\approx 3\times 10^{51}$.  Moreover, we note that the
model is parameterized for the case of non-relativistic ejecta speeds
and is therefore inconsistent with the implied relativistic
velocity.  A consistent, non-relativistic solution to the
\citet{wpm01} model could not be found for the case of SN\,2003L.

\section{Modeling the X-ray emission}
\label{sec:xray_model}

X-ray emission in supernovae can be produced by
three processes \citep{flc96}: (1) non-thermal synchrotron emission
from radiating electrons, (2) thermal (free-free) bremsstrahlung
emission from material in the circumstellar shock and/or the ejecta
reverse shock, and (3) inverse Compton scattering of photospheric
photons by relativistic electrons.  We examine each of
these scenarios in the context of the bright X-ray emission of
SN\,2003L.

\subsection{Synchrotron Emission}
Extrapolating the (optically thin) synchrotron emission at $t\approx
40$ days from the radio to the X-ray band as $f_{\nu} \propto
\nu^{\beta}$ with $\beta\approx -1.1$, we find that the synchrotron
emission under-predicts the observed X-ray flux by a factor of $\sim
50$.  The discrepancy is significantly larger when a synchrotron
cooling break, $\nu_c$, is included, beyond which the spectrum
steepens by $\Delta\beta=-0.5$ (see the Appendix).  Using
Equation~\ref{eqn:nu_c} together with the magnetic field evolution
derived in Section~\ref{sec:ssa}, the cooling frequency at $t\approx
40$ days is $\nu_c \approx 1.1 \times 10^{11}$ Hz.  Including this
synchrotron break and extrapolating from $\nu_c$ to $\nu_{\rm X-ray}$
as $f_{\nu} \propto \nu^{-1.6}$, the synchrotron emission in the X-ray
grossly under-predicts the observed flux by 5 orders of magnitude.  We
also note that the flat spectrum, $\beta\approx -0.5$, of the observed
X-rays is inconsistent with the steep synchrotron spectral index
observed in the radio bands, $\beta\approx -1.1$.  We therefore
conclude that the X-ray emission is not due to synchrotron processes.

\subsection{Thermal Bremsstrahlung Emission}
In a scenario dominated by thermal bremsstrahlung processes, the X-ray
emission is produced by the forward shock plowing into circumstellar
material and/or the reverse shock heating of the ejecta.  In this
case, the strength of the X-ray emission depends of the density of the
emitting material.  The shocked material then cools by free-free
emission processes.  For an $n_e\propto r^{-2}$ wind density profile,
free-free luminosity from the forward or reverse shock is
generalized  \citep{cf01,scb+03} roughly by

\begin{equation}
L_X \approx 7.5\times 10^{34} C_L \left( \frac{\dot{M}}{10^{-5}~\rm M_{\odot}~yr^{-1}}\right)^2 \left( \frac{1000~\rm km/s}{v_w}\right)^2 \left(\frac{t}{40~\rm days}\right)^{-1} ~~\rm erg~s^{-1}
\end{equation}
\label{eqn:L_X}

\noindent
where $C_L$ is a constant such that $C_L=C_{\rm FS}=1$ for the forward
shock and $C_L=C_{\rm RS}=(n-3)(n-4)^2/(4n-8)$ for the reverse shock.
For SSA Model 1 (Section~\ref{sec:ssa}), we find $\alpha_r\approx
0.96$ and thus $n\approx 27$ and $C_{\rm RS}\approx 127$.  Using the
mass loss rate derived from Model 1, the {\it combined} free-free
luminosity from the forward and reverse shocks is therefore $L_{X,\rm
FS+RS}\approx 5.4\times 10^{36}~\rm erg~s^{-1}$, and a factor $\sim
10^3$ fainter than that observed.  On the other hand, if we assume the
X-ray emission {\it must} be dominated by free-free processes, then
this implies a mass loss rate of $2.7\times 10^{-4}~\rm M_{\odot}~
yr^{-1}$ ($v_w=1000~\rm km~s^{-1}$), which is inconsistent with both
the rate predicted by radio SSA models as well as that associated with
FFA modeling (Section~\ref{sec:ffa}).

\subsection{Inverse Compton Cooling}
X-ray emission from supernovae can also be produced by inverse Compton
(IC) cooling of the relativistic electrons by the optical photons as
was claimed to be important for the X-ray emission of SN\,2002ap
\citep{bf04}.  In this scenario, the ratio of energy densities in
magnetic fields, $U_B$, and photons, $U_{\rm ph}$, is given roughly by

\begin{equation}
\frac{f_{\rm radio}}{f_{\rm X-ray}}\approx \frac{U_B}{U_{\rm ph}}
\end{equation} 

\noindent
where $f_{\rm radio}\equiv \nu f_{\nu}$ is the optically thin
synchrotron flux in the radio band and $f_{X}$ is the corresponding
X-ray flux.  At $t\approx 40$ days, the observed radio flux was
$f_{\rm radio}\approx 6.7\times 10^{-16}~\rm erg~cm^{-2}~s^{-1}$ and
the X-ray flux was $f_{\rm X-ray}\approx 9.2\times
10^{-15}\,\rm~erg~cm^{-2}~s^{-1}$, giving a ratio of 
$f_{\rm radio}/f_{\rm X-ray}\approx 0.074$.

Assuming the timescale for Compton cooling is comparable to the epoch
{\it Chandra} observations, $t_{\rm comp} \approx 40$ days, and using
Equations (14--18) of \citet{bf04}, we are able to roughly match the
observed X-ray luminosity.  Since we see no evidence for IC cooling
affects within our radio light-curves, we place a lower limit of
$\nu_{\rm IC} \ge 22$ GHz.  This implies constraints on the magnetic
field and shock velocity of $B\ge 0.4$ G and $v\ge 0.09c$, consistent
with the values predicted from SSA radio analyses
(\S~\ref{sec:ssa}). The corresponding limit on the energy density in
magnetic fields is thus $U_B\gtrsim 0.0048~\rm erg~cm^{-3}$ and is
therefore consistent with the value found through SSA Model 1
(\S~\ref{sec:ssa}). We note that in these estimates, a bolometric
luminosity of $L_{\rm bol, 42}\approx 1.5$ has been very roughly
estimated based on fitting the SN\,2003L optical light-curve
\citep{skg+04}.

While inverse Compton cooling appears to be a feasible process to
produce the luminous X-ray emission of SN\,2003L, it predicts a
steep spectral index, $\beta_{\rm X-ray}\approx \beta_{\rm
radio}\approx -1.1$, which is inconsistent with the observed value of
$\beta\approx -0.5$ ($2-\sigma$).  We note, however, that the poor photon
statistics may be responsible for this discrepancy.

\section{Discussion and Conclusions}
\label{sec:disc}
As the first Type Ibc supernova with a radio and X-ray luminosity
within an order of magnitude of SN\,1998bw, SN\,2003L is clearly an
unusual event.  Here, we summarize our findings.  Both SSA and FFA models
(Sections~\ref{sec:ssa} and \ref{sec:ffa}) show that the energy of the
ejecta is $\gtrsim 10^{48}$ erg and nearly comparable to that of
SN\,1998bw at a similar epoch.  In Case 1 the equipartition and SSA
analyses predict a sub-relativistic velocity, $\overline{v}\approx 0.2c$,
and a relatively low mass loss rate, $\dot{M}\approx 7.5\times
10^{-6}~M_{\odot}~\rm yr^{-1}$ which are roughly consistent with a
scenario where inverse Compton scattering produces the observed X-ray
emission.  In Case 2, the complex FFA models reproduce the observed radio
light-curves well, yet they imply a highly relativistic ejecta
speed, $\overline{v}\approx 14c$, or a huge energy output of $E\ge
10^{51}$ erg, as well as a high mass loss rate of $\dot{M}\approx
10^{-2}~M_{\odot}~\rm yr^{-1}$ which over-predicts the observed 
X-ray emission assuming a thermal bremsstrahlung emission model.

Our VLBA observation at $t\approx 65$ days implies an ejecta size
which is inconsistent with the extreme velocities required by the
complex FFA model in Case 2. Only by deviating far from equipartition and
significantly increasing the total energy of the ejecta can this
inconsistency be resolved.  Moreover, the predicted mass loss rate is
at least two orders of magnitude above the observed rates from
Wolf-Rayet stars.  In conclusion, we find Case 2 to be infeasible and
we therefore adopt Case 1 the most likely scenario. We note that Case 1
predicts a mass loss rate comparable to the low end of values observed
for local Wolf-Rayet stars.

While similar to SN\,1998bw in radio luminosity and ejecta energy, a
main distinguishing characteristic between SN\,2003L and SN\,1998bw is
their shock velocity: while the radio ejecta of SN\,1998bw attained
mildly-relativistic speeds, that of SN\,2003L was clearly
sub-relativistic as shown in Section~\ref{sec:ssa}.  This discrepancy
is echoed by optical spectroscopy which showed photospheric velocities
in excess of $3\times 10^4~\rm km~s^{-1}$ for SN\,1998bw
\citep{pcd+01} and $10^{4}~\rm km~s^{-1}$ for SN\,2003L
\citep{vcd+03,mck+03,skg+04}.  Therefore, we see that for SN\,1998bw,
mildly-relativistic ejecta was coupled with bright radio emission,
while for SN\,2003L the emission is bright {\it despite} the modest
ejecta velocities.  The question is thus, why is the radio ejecta of
SN\,2003L especially luminous and energetic?

One hypothesis considers the distribution of kinetic energy according
to ejecta velocity.  SN\,1998bw is a clear case where a significant
fraction of the explosion energy has been deposited into
mildly-relativistic material.  For SN\,2003L, however, the bulk of the
energy is coupled to sub-relativistic ejecta.  \citet{tmm01} have
shown that in some cases, high velocity ejecta can be produced from
a whip effect racing down the ejecta.  The observed diversity in
the coupling between velocity and energy could therefore be intrinsic
to the SN Ibc population. A similar situation is observed for
gamma-ray bursts which partition their energy into ultra-relativistic
emission ($\gamma$-rays) and mildly-relativistic afterglow emission
(X-rays, optical, radio) by variable fractions.

Through our ongoing radio surveys of local SNe Ibc and cosmological
gamma-ray bursts, we continue to map out the energetics of these
explosions.  Our goal is to understand the link between Type Ibc
supernovae and GRBs by studying properties of the explosion (ejecta
velocity and kinetic energy) in addition to those of the progenitor
star (CSM density profile and evolutionary mass loss rate).  Through
these further studies, we aim to shed light on the illusive bridge
between these two classes of cosmic explosions.

\acknowledgments We are indebted to Barry Clark for his generous
scheduling of the VLA for this project, and to Jim Ulvestad for 
enabling the VLBA observations. The authors thank Re'em Sari,
Claes Fransson, Eli Waxman, Dick Sramek and Michael Rupen for useful
discussions.  Caltech SN research is supported by NSF and NASA
grants.  AMS acknowledges support by the National Radio Astronomy
Observatory Graduate Summer Student Research Assistantship program.
Research by EB is supported by NASA through Hubble Fellowship grant
HST-HF-01171.01 awarded by the Space Telescope Science Institute,
which is operated by the Association of Universities for Research in
Astronomy, Inc., for NASA, under contract NAS 5-26555.  RAC thanks NSF
grant AST-0307366.


\clearpage

\appendix

\centerline{\large \bf Appendix \rm \normalsize}

\section{The Synchrotron Self-Absorption Model}
Here we describe the SSA model prescription applied in
Section~\ref{sec:ssa}.  The model was adapted from the formalism of
\citep{fwk00}, designed to describe gamma-ray burst radio emission
following the transition to sub-relativistic, adiabatic expansion.  We
replace the assumption of Sedov-Neumann-Taylor (SNT) dynamics
\citep{zr02} with a general parameterization of the shock evolution.
This substitution enables us to study the early supernova synchrotron
emission, while the ejecta is in the free-expansion phase.  We note
that by adopting the scalings given by the SNT dynamics into our
generalized equations, the formalism of \citet{fwk00} is fully
recovered.

\subsection{Physical Assumptions}
\label{sec:app_phys_assump}

To first order, we assume that the supernova ejecta is undergoing
spherical, homologous expansion at sub-relativistic velocity, $v$.  At
any time, $t$, the observed synchrotron emission originates from a
thin shell of radiating electrons with radius, $r$, and thickness,
$r/\eta$ (with $\eta \approx 10$).  The electrons are accelerated into
a power-law energy distribution, $N(\gamma)\propto \gamma^{-p}$, above
a minimum Lorentz factor, $\gamma_m$.  We adopt the standard
assumption that the energy density of the ejecta is partitioned
between the fraction in relativistic electrons, $\epsilon_e$, and the
fraction in magnetic fields, $\epsilon_B$.  For simplicity, we denote
the ratio of these energies as $\frak{F}\equiv \epsilon_e/\epsilon_B$
and further assume that it evolves as $\frak{F} \propto
t^{\alpha_{\frak{F}}}$.

The temporal evolution of the radius, velocity, minimum Lorentz factor
and magnetic field are then parameterized as

\begin{eqnarray}
r & = & r_0\left(\frac{t}{t_0}\right)^{\alpha_r} 
\label{eqn:r_scaling} \\
v & = & v_0\left(\frac{t}{t_0}\right)^{\alpha_r-1}
\label{eqn:v_scaling} \\
\gamma_m & = & \gamma_{m,0}   \left(\frac{t}{t_0}\right)^{\alpha_{\gamma}} 
\label{eqn:gamma_scaling} \\
B & = & B_0\left(\frac{t}{t_0}\right)^{\alpha_B}
\label{eqn:B_scaling}
\end{eqnarray}

\noindent
where the subscript $0$ corresponds to the parameter values at an
(arbitrary) reference time, $t_0$.  Here, the indices $\alpha_r$,
$\alpha_{\gamma}$, $\alpha_B$ and $\alpha_{\frak{F}}$ are determined
by the hydrodynamic evolution of the ejecta, and are constrained
according to the assumed evolutionary model (see
Section~\ref{sec:hydro}).

\subsection{Emission Spectrum}

In characterizing the synchrotron emission spectrum, we use the
standard formalism of  \citet{rl79}.  Within this framework, the
synchrotron power per unit frequency emitted by a single electron is
given by

\begin{equation}
P(\nu,~\gamma)=\frac{e^3 B}{m_e c^2} F\left(\frac{\nu}{\nu_{\rm crit} (B,~\gamma)}\right)
\label{eqn:power}
\end{equation}

\noindent
where $\gamma$, $e$ and $m_e$ are the Lorentz factor, charge and mass
of the electron, respectively.  The critical frequency, $\nu_{\rm
crit}$, is defined as

\begin{equation}
\nu_{\rm crit}\equiv \gamma^2 \left( \frac{e B}{2\pi m_e c}\right)
\label{eqn:nu_crit}
\end{equation}

\noindent
 \citep{rl79}.  Adopting the notation $x\equiv (2/3)(\nu/\nu_{\rm crit})$, the
function $F(x)$ describes the total synchrotron power spectrum.

\begin{equation}
F(x)\equiv x \int_x^{\infty} K_{5/3}(\xi )d\xi
\label{eqn:F_x}
\end{equation}

\noindent
where $K_{5/3}$ is the modified Bessel function of $2/3$ order
 \citep{rl79}.  This function peaks at $x=0.29$ and
decays rapidly for $x > 1$. 

Applying the temporal scalings for $r$, $\beta$, $B$ and $\gamma_m$
(Equations~\ref{eqn:r_scaling} - \ref{eqn:B_scaling}), the flux density
from a uniform shell of radiating electrons is then given by

\begin{equation}
f_{\nu}(t)=C_f\left(\frac{t}{t_0}\right)^{(4\alpha_r-\alpha_B)/2} [(1-e^{-\tau_{\nu}^{\zeta}(t)})]^{1/\zeta} \nu^{5/2} F_3(x) F_2^{-1}(x)~~\rm erg/s/Hz/cm^{2}
\label{eqn:flux}
\end{equation}

\noindent
where the optical depth, $\tau_{\nu}(t)$, is defined

\begin{equation}
\tau_{\nu}(t)=C_{\tau}\left(\frac{t}{t_0}\right)^{(p-2)\alpha_{\gamma}+(3+p/2)\alpha_B+\alpha_r+\alpha_{\frak{F}}} \nu^{-(p+4)/2} F_2(x)
\label{eqn:tau}
\end{equation}

\noindent
and $\zeta=[0,1]$ parameterizes the sharpness of the spectral break between
optically thick and thin regimes.  We adopt
$\nu_m\equiv \nu_{\rm crit}(\gamma=\gamma_m)$ as the characteristic
synchrotron frequency,

\begin{equation}
\nu_m(t)=\nu_{m,0} \left(\frac{t}{t_0}\right)^{2\alpha_{\gamma}+\alpha_B}~~\rm Hz,
\label{eqn:nu_m}
\end{equation}

\noindent
and for $x=(2/3)(\nu/\nu_m(t))$, the functions, $F_2(x)$ and $F_3(x)$,
are defined

\begin{equation}
F_2(x)\equiv \sqrt{3}\int_0^{x} F(y) y^{(p-2)/2} dy,~~~F_3(x)\equiv \sqrt{3}\int_0^{x} F(y) y^{(p-3)/2} dy
\label{eqn:F_2}
\end{equation}

\noindent 
where $y$ is a simple integration variable representing the range
$y=[0,x]$.  The temporal dependencies of $F_2(x)$ and 
$F_3(x)$ can be computed numerically.  For $\nu \ll \nu_m$, we find the 
following scalings

\begin{equation}
F_2(x) \propto t^{0.49(p+0.73)},~~~F_3(x) \propto t^{0.47(p-0.14)}
\label{eqn:F2_scaling}
\end{equation}

\noindent
and for $\nu \gg \nu_m$, neither function evolves.
Therefore, using equations \ref{eqn:flux} through \ref{eqn:F2_scaling}, the
model flux density at any time is strictly determined by three
parameters: $C_f$, $C_{\tau}$ and $\nu_{m,0}$:

\begin{eqnarray}
C_f & \equiv &\frac{2\pi}{2+p} m_e \left(\frac{r_0}{d}\right)^2 \left(\frac{2\pi m_e c}{e B_0}\right)^{1/2}
\label{eqn:C_f} \\
C_{\tau} & \equiv & \frac{(p+2)(p-2)}{4\pi \eta} \gamma_{m,0}^{(p-1)} \left(\frac{B_0^2}{8\pi}  \frak{F}_0 \right) \left(\frac{e^3 B_0 r_0}{m_e^3 c^4 \gamma_{m,0}}\right) \left( \frac{e B_0}{2\pi m_e c}\right)^{p/2} 
\label{eqn:C_tau} \\
\nu_{m,0} & \equiv & \frac{1}{2\pi} \gamma_{m,0}^2 \frac{e B_0}{m_e c}
\label{eqn:nu_m_0}
\end{eqnarray}

\noindent
where $d$ is the distance to the source and $\frak{F}_0$ is the value
of $\frak{F}$ at $t_0$.  Equations~\ref{eqn:C_f} - \ref{eqn:nu_m_0}
above show that $C_f$, $C_{\tau}$ and $\nu_{m,0}$ are in turn
determined by four physical parameters: $B_0$, $r_0$, $\gamma_{m,0}$
and $\frak{F}_0$.

\subsection{Temporal and Spectral Evolution}
\label{sec:scalings}

We use the above formalism to determine the spectral evolution of the
synchrotron emission for a given ordering of the synchrotron break
frequencies, $\nu_a$ and $\nu_m$.  Here, we define the
self-absorption frequency, $\nu_a$, as the frequency at which the
optical depth is unity: $\tau_{\nu_a}\equiv \tau (\nu=\nu_a)=1$.  

For completeness, we give the temporal and spectral evolution for
higher frequencies near the synchrotron cooling frequency, $\nu_c$,
defined as the the frequency above which electrons cool efficiently.
This frequency is given by

\begin{equation}
\nu_c = \frac{18\pi m_e c e}{t^2 \sigma_T^2 B^3}~~\rm Hz
\label{eqn:nu_c}
\end{equation}

\noindent
and is typically located between the radio and optical observing
bands during the first few years after the supernova explosion.  

\subsubsection{Case 1: $\nu_a \ll \nu_m$}
In Case 1, the self-absorption frequency is well below the
characteristic synchrotron frequency,  $\nu_a \ll \nu_m$.
In this scenario, the spectrum peaks at $\nu_{p}\approx \nu_m$.
It can be shown that the temporal scalings associated with the peak are
then given by,

\begin{equation}
\nu_{p} \approx \nu_m\propto t^{2\alpha_{\gamma}+\alpha_B},~~~
f_{\nu_{p}}\approx  f_{\nu_m}\propto t^{3\alpha_r+3\alpha_B-\alpha_{\gamma}+\alpha_{\frak{F}}}
\label{eqn:peak_1}
\end{equation}

\noindent
Here, $\nu_a \ll \nu_m$ so the evolution of $\nu_a$ and $f_{\nu_a}$
depend on the temporal scaling of $F_2(x)$
(Equation~\ref{eqn:F2_scaling}).  We find

\begin{equation}
\nu_a\propto t^{(2(p-136)\alpha_{\gamma}+(p+264)\alpha_B+100\alpha_r+100\alpha_{\frak{F}})/(p+164)},~~~
f_{\nu_a}\propto t^{(9\alpha_r+8\alpha_B-5\alpha_{\gamma}+\alpha_{\nu_a}+3\alpha_{\frak{F}})/3}
\label{eqn:nu_a_1}
\end{equation}

\noindent
where, for simplicity, we use $\alpha_{\nu_a}$ to denote the temporal
dependence of $\nu_a$.  For all frequencies, the temporal and frequency
dependence of $f_{\nu}$ is then generalized by

\begin{equation}
f_{\nu} \propto 
  \begin{cases}
    \nu^2 t^{(9\alpha_r+8\alpha_B-5\alpha_{\gamma}-5\alpha_{\nu_a}+3\alpha_{\frak{F}})/3} & \nu < \nu_a \\
    \nu^{1/3} t^{(9\alpha_r+8\alpha_B-5\alpha_{\gamma}+3\alpha_{\frak{F}})/3} & \nu_a < \nu < \nu_m \\
    \nu^{-(p-1)/2} t^{(6\alpha_r+(5+p)\alpha_B+2(p-2)\alpha_{\gamma} +2\alpha_{\frak{F}})/2} & \nu_m < \nu < \nu_c \\
    \nu^{-p/2} t^{(6\alpha_r+(8-5p)\alpha_B+2(p-2)\alpha_{\gamma}+2\alpha_{\frak{F}}-4p+2)/2} & \nu_c < \nu
  \end{cases}
\label{eqn:case_1}
\end{equation}

\noindent
where the temporal scalings associated with the synchrotron cooling frequency
are

\begin{equation}
\nu_c\propto t^{-3\alpha_B-2},~~~
f_{\nu_c}\propto t^{3\alpha_r+(4-p)\alpha_B+(p-2)\alpha_{\gamma}+\alpha_{\frak{F}}-p+1}.
\label{eqn:nu_c_1}
\end{equation}

\subsubsection{Case 2: $\nu_m \ll \nu_a$}
In  Case 2,  the  characteristic synchrotron  frequency  is below  the
self-absorption  frequency,  $\nu_m \ll  \nu_a$.   In  this case,  the
spectral  peak occurs at  $\nu_{p}\approx  \nu_a$.  Since $F_2(x)$
is constant  for $\nu \gg \nu_m$, the scalings simplify to 

\begin{equation}
\nu_{p}\approx \nu_a\propto t^{(2(p-2)\alpha_{\gamma}+2(3+p/2)\alpha_B+2\alpha_r+2\alpha_{\frak{F}})/(p+4)},~~~
f_{\nu_{p}}\approx f_{\nu_a}\propto t^{(4\alpha_r-\alpha_B+5\alpha_{\nu_a})/2}.
\label{eqn:peak_2}
\end{equation}

\noindent
In this scenario, the evolution of $\nu_m$ is given by
Equation~\ref{eqn:nu_m} while $F_{\nu_m}$ scales as

\begin{equation}
f_{\nu_m} \propto t^{2\alpha_r+2\alpha_B+5\alpha_{\gamma}}.
\end{equation}
\label{eqn:f_nu_m_2}

\noindent
For all frequencies, the temporal and frequency dependence of $f_{\nu}$
is then generalized by

\begin{equation}
f_{\nu} \propto 
  \begin{cases}
    \nu^2 t^{2\alpha_r+\alpha_{\gamma}} & \nu < \nu_m \\
    \nu^{1/3} t^{(4\alpha_r-\alpha_B)/2} & \nu_m < \nu < \nu_a \\
    \nu^{-(p-1)/2} t^{(4\alpha_r-\alpha_B+(4+p)\alpha_{\nu_a})/2} & \nu_a < \nu < \nu_c \\
    \nu^{-p/2} t^{(2\alpha_r+(1-3p)\alpha_B+(2+p/2)\alpha_{\nu_a}-2p+1)} & \nu_c < \nu
  \end{cases}
\end{equation}
\label{eqn:case_2}

\noindent
where the temporal scalings associated with the synchrotron cooling frequency
are

\begin{equation}
\nu_c\propto t^{-3\alpha_B-2},~~~
f_{\nu_c}\propto t^{(4\alpha_r+(3-3p)\alpha_B+(4+p)\alpha_{\nu_a}-2p+2)/2}.
\label{eqn:nu_c_2}
\end{equation}

\bibliographystyle{apj1b} 

\clearpage

\begin{deluxetable}{lrrrrrr}
\tablecaption{VLA radio flux density measurements of SN\,2003L}
\label{tab:vla}
\tablewidth{0pt}
\tablehead{ \colhead{Date} & \colhead{$\Delta t$} & \colhead{$F_{4.9\rm~GHz}\pm \sigma$} & \colhead{$F_{8.5\rm~GHz}\pm \sigma$} & \colhead{$F_{15.0\rm~GHz}\pm \sigma$} & \colhead{$F_{22.5\rm~GHz}\pm \sigma$} & \colhead{Array} \\
\colhead{(UT)} & \colhead{(days)} & \colhead{($\mu$Jy)} & \colhead{($\mu$Jy)} & \colhead{($\mu$Jy)} & \colhead{($\mu$Jy)} & \colhead{Config.} \\
}
\startdata
2003 Jan 26.2  &  25.2 &   \nodata    &  743$\pm$39 &      \nodata &     \nodata  &  DnC \\
2003 Jan 28.3  &  27.3 &   \nodata    &  810$\pm$63 &      \nodata & 3179$\pm$85  &  DnC \\
2003 Jan 29.3  &  28.3 &   \nodata    &  848$\pm$65 &      \nodata & 3119$\pm$86  &  DnC \\
2003 Jan 30.4  &  29.4 &   \nodata    &  966$\pm$60 &      \nodata & 3122$\pm$75  &  DnC \\
2003 Jan 31.3  &  30.3 &   \nodata    & 1051$\pm$62 &      \nodata & 3080$\pm$86  &  DnC \\
2003 Feb 1.2  &  31.2 &   \nodata    &  947$\pm$52 &      \nodata & 3252$\pm$70  &  DnC \\
2003 Feb 2.3  &  32.3 &   \nodata    &  883$\pm$53 &      \nodata & 2973$\pm$99  &  DnC \\
2003 Feb 3.3  &  33.3 &   \nodata    &  \nodata    & 2287$\pm$121 &     \nodata  &  DnC \\
2003 Feb 6.3  &  36.3 &   \nodata    & 1147$\pm$47 &      \nodata &     \nodata  &  DnC \\
2003 Feb 8.3  &  38.3 &   \nodata    & 1211$\pm$39 &      \nodata &     \nodata  &  D   \\
2003 Feb 11.6 &  41.6 &   \nodata    & 1483$\pm$76 & 2686$\pm$156 & 2956$\pm$96  &  D   \\
2003 Feb 14.5 &  44.5 &   \nodata    & 1448$\pm$62 & 2753$\pm$153 & 2684$\pm$124 &  D   \\
2003 Feb 16.4 &  46.4 &   \nodata    & 1413$\pm$55 & 2719$\pm$144 &      \nodata &  D   \\
2003 Feb 18.4 &  48.4 &   \nodata    & 1546$\pm$45 &      \nodata &      \nodata &  D   \\
2003 Feb 22.3 &  52.3 &   \nodata    &    \nodata  & 3014$\pm$137 & 2714$\pm$68  &  D   \\
2003 Feb 23.3 &  53.3 &   \nodata    & 1854$\pm$41 &      \nodata &      \nodata &  D   \\
2003 Feb 25.3 &  55.3 &   \nodata    & 1969$\pm$53 & 2799$\pm$149 & 2838$\pm$99  &  D   \\
2003 Feb 26.3 &  56.3 &   \nodata    &    \nodata  & 2865$\pm$131 &      \nodata &  D   \\ 
2003 Mar 2.3     &  60.3 &   \nodata    & 2283$\pm$51 & 2924$\pm$132 &      \nodata &  D   \\
2003 Mar 7.3     &  65.3 &   848$\pm$64\tablenotemark{a}    &    \nodata  &      \nodata & 2263$\pm$73  &  D   \\
2003 Mar 8.4     &  66.4 &   \nodata    & 2351$\pm$47 & 2857$\pm$123 & 2486$\pm$66  &  D   \\
2003 Mar 10.3    &  68.3 &   \nodata    & 2081$\pm$52 & 2966$\pm$115 &      \nodata &  D   \\
2003 Mar 14.4    &  72.4 &   \nodata    & 2527$\pm$52 & 2901$\pm$150 & 2422$\pm$70  &  D   \\
2003 Mar 17.4    &  75.4 &   \nodata    & 2673$\pm$57 & 3055$\pm$173 &      \nodata &  D   \\
2003 Mar 20.3    &  78.3 &   \nodata    & 2422$\pm$61 &      \nodata & 2130$\pm$223 &  D   \\
2003 Mar 23.3    &  81.3 &   \nodata    & 2500$\pm$43 & 2664$\pm$118 &      \nodata &  D   \\
2003 Mar 27.4    &  85.4 &   \nodata    & 2776$\pm$51 &      \nodata & 1833$\pm$100 &  D   \\
2003 Apr 1.2     &  90.2 &   \nodata    & 2551$\pm$54 &      \nodata &      \nodata &  D   \\
2003 Apr 2.1     &  91.1 &   \nodata    &     \nodata &      \nodata & 1819$\pm$141 &  D   \\
2003 Apr 10.4    &  99.4 &   \nodata    &     \nodata & 2581$\pm$320 &\nodata &  D   \\
2003 Apr 30.0    & 119.0 &   \nodata    & 2532$\pm$75 & \nodata &  1078$\pm$173& D \\
2003 May 3.1       & 122.1 &   \nodata    & 2534$\pm$80 & \nodata & 1327$\pm$114  &  D \\
2003 May 16.3      & 135.3 &   \nodata    & 2455$\pm$96 & 1925$\pm$282 & 1146$\pm$165  &  D \\
2003 May 28.1      & 147.0 & 1989$\pm$41  & 2200$\pm$47 &  1452$\pm$235 & 941$\pm$161 &  D   \\
2003 Jun 4.0      & 154.0 & 2410$\pm$114 & 2151$\pm$77 & \nodata &  \nodata &  A   \\
2003 Jun 17.1     & 167.1 & 2555$\pm$83  & 2210$\pm$68 & 1279$\pm$226 &      \nodata &  A   \\
2003 Jul 1.1      & 181.1 &   \nodata    & 2262$\pm$63 &      \nodata &      \nodata &  A   \\
2003 Jul 9.0      & 189.0 & 2485$\pm$90  & 2075$\pm$61 & 1135$\pm$125 &      \nodata &  A   \\
2003 Jul 24.0     & 204.0 & 2380$\pm$81  & 2048$\pm$66 & 1177$\pm$166 &      \nodata &  A   \\
2003 Aug 11.9   & 222.9 & \nodata & \nodata & \nodata &  670$\pm$148 & A \\
2003 Aug 15.8   & 226.8 & 2346$\pm$84  & 1646$\pm$74 &  824$\pm$237 & \nodata &  A   \\ 
2003 Sep 15.7 & 257.7 & 2177$\pm$87 & 1151$\pm$53 &  768$\pm$165 & \nodata &  A   \\
2003 Oct 6.7   & 278.6 & 2140$\pm$40  & 1244$\pm$58 &      \nodata &      \nodata &  BnA \\ 
2003 Oct 11.7  & 283.7 & 2177$\pm$79  & 1176$\pm$45 &  612$\pm$122 &      \nodata &  BnA \\
2003 Oct 27.7  & 299.7 & 1937$\pm$70  & 1200$\pm$42 &  520$\pm$124 &      \nodata &  B   \\
2003 Nov 1.7  & 304.7 &    \nodata   & 1115$\pm$55 &      \nodata &      \nodata &  B   \\
2003 Nov 14.6 & 317.6 & 1713$\pm$73  & 1075$\pm$52 &      \nodata &      \nodata &  B   \\
2003 Nov 25.5 & 328.5 & 1689$\pm$67  &  981$\pm$49 &  148$\pm$148 &      \nodata &  B   \\
2003 Dec 6.6  & 339.6 & 1980$\pm$69  &  961$\pm$42 &  506$\pm$149&      \nodata &  B   \\
2004 Jan 26.4  & 390.4 &    \nodata   &  714$\pm$38 &      \nodata &      \nodata &  B   \\
2004 Feb 8.3  & 403.3 &    \nodata   &  863$\pm$42 &      \nodata &      \nodata &  CnB \\
2004 Mar 6.4     & 430.4 & 1281$\pm$42  &     \nodata &      \nodata &      \nodata &  C   \\
2004 Mar 20.2    & 444.2 &    \nodata   &  668$\pm$30 &      \nodata &      \nodata &  C   \\
2004 Apr 21.0    & 476.0 &    \nodata   &     \nodata &  125$\pm$137 &      \nodata &  C   \\
2004 May 21.0   & 506.0 &    \nodata    & \nodata   & \nodata  & 280$\pm$53 & D \\
2004 Jul 27.0   & 573.0 &    \nodata   &  592$\pm$48 &      \nodata &      \nodata &  D   \\
\enddata
\tablenotetext{a}{VLBA observation}
\label{tab:vla}
\end{deluxetable}

\clearpage

\clearpage

\clearpage

\clearpage

\clearpage

\clearpage

\clearpage

\end{document}